\documentclass[journal,transmag]{IEEEtran}
\usepackage{amsmath}
\usepackage{bm}
\usepackage{amsfonts}

\usepackage[font=footnotesize]{caption}
\usepackage[font=footnotesize]{subcaption}
\usepackage{stmaryrd}
\usepackage{pgfplots}
\usepackage{siunitx}

\usepackage{hyperref}
\usepackage{color}
\usepackage{breqn}
\usepackage{multirow}
\usepackage[capitalise]{cleveref}
\crefname{equation}{}{}\usepackage[top=0.7in, left=0.65in]{geometry}
\usepackage{graphicx}
\usepackage{xargs} \usepackage{pdfcomment}
\usepackage{tikz}
\usepackage{pgfplots}\usetikzlibrary{calc,spy,shapes,arrows, patterns}
\usepackage{contour}
\usepackage[normalem]{ulem}

\usepackage{eso-pic}

\DeclareMathOperator{\thd}{THD}
\DeclareMathOperator{\setI}{\mathcal{I}}
\DeclareMathOperator{\setN}{\mathbb{N}}

\DeclareMathOperator{\x}{\mathbf{x}}

\AddToShipoutPicture*{\footnotesize\sffamily\raisebox{1cm}{\hspace{1.65cm}\fbox{\parbox{\textwidth}{\copyright~2020
IEEE. Personal use of this material is permitted. Permission from
IEEE must be obtained for all other uses, in any current or future
media, including reprinting/republishing this material for
advertising or promotional purposes, creating new collective works,
for resale or redistribution to servers or lists, or reuse of any
copyrighted component of this work in other works.}}}}
\begin{document}
\title{Shape Optimization of Rotating Electric Machines using Isogeometric Analysis and Harmonic Stator-Rotor Coupling}
\author{\IEEEauthorblockN{
Melina Merkel\IEEEauthorrefmark{1,2},
Peter Gangl\IEEEauthorrefmark{3} and
Sebastian Sch\"ops\IEEEauthorrefmark{1,2}}
\IEEEauthorblockA{\IEEEauthorrefmark{1}Institut f\"ur Teilchenbeschleunigung und Elektromagnetische Felder (TEMF), Technische Universit\"at Darmstadt, Germany}
\IEEEauthorblockA{\IEEEauthorrefmark{2}Centre for Computational Engineering, Technische Universit\"at Darmstadt, Germany}
\IEEEauthorblockA{\IEEEauthorrefmark{3}Institut f\"ur Angewandte Mathematik, Technische Universit\"at Graz, Austria}}

\IEEEtitleabstractindextext{\begin{abstract}
This work deals with shape optimization of electric machines using isogeometric analysis. Isogeometric analysis is particularly well suited for shape optimization as it allows to easily modify the geometry without remeshing the domain. A 6-pole permanent magnet synchronous machine is modeled using a multipatch isogeometric approach and rotation of the machine is realized by modeling the stator and rotor domain separately and coupling them at the interface using harmonic basis functions. Shape optimization is applied to the model minimizing 
the total harmonic distortion of the electromotive force as a goal functional.
\end{abstract}

}

\maketitle
\thispagestyle{empty}
\pagestyle{empty} 

\section{Introduction}
As a result of the energy transition, the simulation of electromechanical energy converters, in particular electric machines, is becoming increasingly important to obtain  efficient and robust designs.
Commonly, a workflow based on analytical estimates and the Finite Element Method (FEM) is used for 2D and finally 3D domains.
In most classical approaches, the geometry is only approximated, e.g., with an accuracy that depends on the mesh refinement. 
These errors can be avoided when using Isogeometric Analysis (IGA) \cite{Hughes_2005aa}, \cite{Buffa_2010aa}, \cite{Bontinck_2017ag}, \cite{Dolz_2018aa} which uses B-splines and/or Non-Uniform Rational B-splines (NURBS) as basis for geometry and solution space. In IGA, the geometry can be easily and smoothly transformed by moving the control points of the splines such that there is no need to remesh the domain when the geometry is modified. This makes IGA very well suited for shape optimization \cite{1809.03377}. 
Numerical optimization based on magnetic equivalent circuits or finite element models has lead to large improvements in the designs of technical applications, e.g., of permanent magnet synchronous machines. During the last 30 years there has been a lot of research on finite element based optimization methods (see, e.g., \cite{Di-Barba_2010aa}, \cite{Duan_2013aa} and the references therein). Originally, mainly gradient-based optimization methods were used (see, e.g., \cite{Russenschuck_1990aa,Weeber_1992aa,Takorabet_1997aa}) but stochastic optimization algorithms became more popular during the last 20 years (see, e.g., \cite{Hameyer_1993aa}, \cite{Lok_2017aa}). Most of the proposed algorithms use stochastic or population-based optimization, e.g., genetic algorithms and particle swarm optimization (see, e.g., \cite{Ma_2015aa}) which have also been extended to multi-objective optimization problems (see, e.g., \cite{Baumgartner_2004aa,Ho_2005aa}). For permanent magnet synchronous machines, stochastic optimization methods are commonly used (see, e.g.,\cite{Cassimere_2009aa,Bash_2011aa,Sizov_2013aa}).

This contribution deals for the first time with shape optimization of a rotating electrical machine discretized with IGA. We use shape calculus to obtain shape derivatives such that gradient-based optimization becomes feasible. The rotor and stator domains are discretized separately and are coupled using harmonic stator-rotor coupling \cite{Bontinck_2018ac}.

The paper is structured as follows: In Section \ref{sec:model} the model of the electric machine and its quantities of interest are introduced.
We introduce the shape optimization problem and derive the formula of the shape derivative using shape calculus in Section \ref{sec:opti}.
In Section \ref{sec:discretization} the discretization of the machine using harmonic stator-rotor coupling and isogeometric analysis is explained and the resulting saddle-point problem is presented.
We conclude the paper by explaining our gradient-based shape optimization algorithm and present numerical results for the minimization of the total harmonic distortion of the electromotive force in Section \ref{sec:num_shape_opti}.

\section{Model of the Electric Machine} \label{sec:model}
Electromagnetic phenomena are described by Maxwell's equations. For many applications it is sufficient to consider the magnetostatic approximation in a working point, i.e., neglecting displacement and eddy currents and nonlinearity of the materials.
In a domain $D_{\mathrm{3D}}$ the magnetostatic formulation of Maxwell's equations is given by
\begin{align}
\nabla\times\left(\nu\nabla \times \mathbf{A}\right) = \mathbf{J}_{\mathrm{src}} + \nabla \times \mathbf{M} , \label{eq:curlcurl}
\end{align}
with the piecewise constant reluctivity $\nu$, the magnetic vector potential $\mathbf{A}$, the current density $\mathbf{J}_{\mathrm{src}}=\sum_k\pmb{\chi}_ki_k$ given by winding functions $\pmb{\chi}_k$ and currents $i_k$ \cite{Schops_2013aa}, the magnetization of the permanent magnets $\mathbf{M}$ and homogeneous Dirichlet boundary conditions $\mathbf{A}\times\mathbf{n}=0$ on $\partial D_{\mathrm{3D}}$, where $\mathbf{n}$ is the normal vector. The source current density $\mathbf{J}_{\mathrm{src}}$ and the permanent magnetization $\mathbf{M}$ vanish outside the coil (${D}_{\mathrm{c}}$) and permanent magnet (${D}_{\mathrm{pm}}$) regions, respectively, see Fig.~\ref{fig:kartoffel}.
\begin{figure}
	\centering
	\begin{tikzpicture}[scale=0.8]
	
	\path [pattern color=red, pattern=north east lines, loosely dashed] plot [smooth cycle, tension=2] coordinates {(2,1.5) (2,2.5) (1,2.7)};
	\draw [black] plot [smooth cycle, tension=2] coordinates {(2,1.5) (2,2.5) (1,2.7)};
	
	\path [pattern color=red, pattern=north east lines, loosely dashed] plot [smooth cycle, tension=2] coordinates {(3,3) (3.5,3.5) (4,2.5)};
	\draw [black] plot [smooth cycle, tension=2] coordinates {(3,3) (3.5,3.5) (4,2.5)};
	
	\path [pattern color=orange, pattern=north east lines, loosely dashed] plot [smooth cycle, tension=2] coordinates {(6,3) (6,4) (4.5,2.5)};
	\draw [black] plot [smooth cycle, tension=2] coordinates {(6,3) (6,4) (4.5,2.5)};
	
    \path [pattern color=blue, pattern=north west lines, dotted] plot [smooth cycle, tension=2, pattern color=blue, pattern=north east lines] coordinates {(0,1) (4,4.5) (6,2)};
	\draw [black] plot [smooth cycle, tension=2] coordinates {(0,1) (4,4.5) (6,2)};
	
	\node at (3.5,1.5) {\contour{blue!20}{${D}$}};
	\node [inner sep=1pt] (dc) at (2.7,2.5) {\contour{red!30}{${D}_{\mathrm{c}}$}};
	\node at (5.5,3.2) {\contour{orange!40}{${D}_{\mathrm{pm}}$}};
	\node at (1.3,4) {$\partial{D}$};
	\draw [->] (dc) -- (3.5,3);
	\draw [->] (dc) -- (1.8,2.2);
	\end{tikzpicture}
\caption{\label{fig:kartoffel}Example for a domain ${D}$ with coil region ${D}_{\mathrm{c}}$, permanent magnet region ${D}_{\mathrm{pm}}$ and domain boundary $\partial {D}$.}
\end{figure}
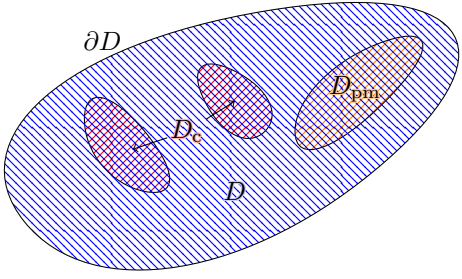
\begin{figure}
  \centering
  \includegraphics[width=0.35\textwidth]{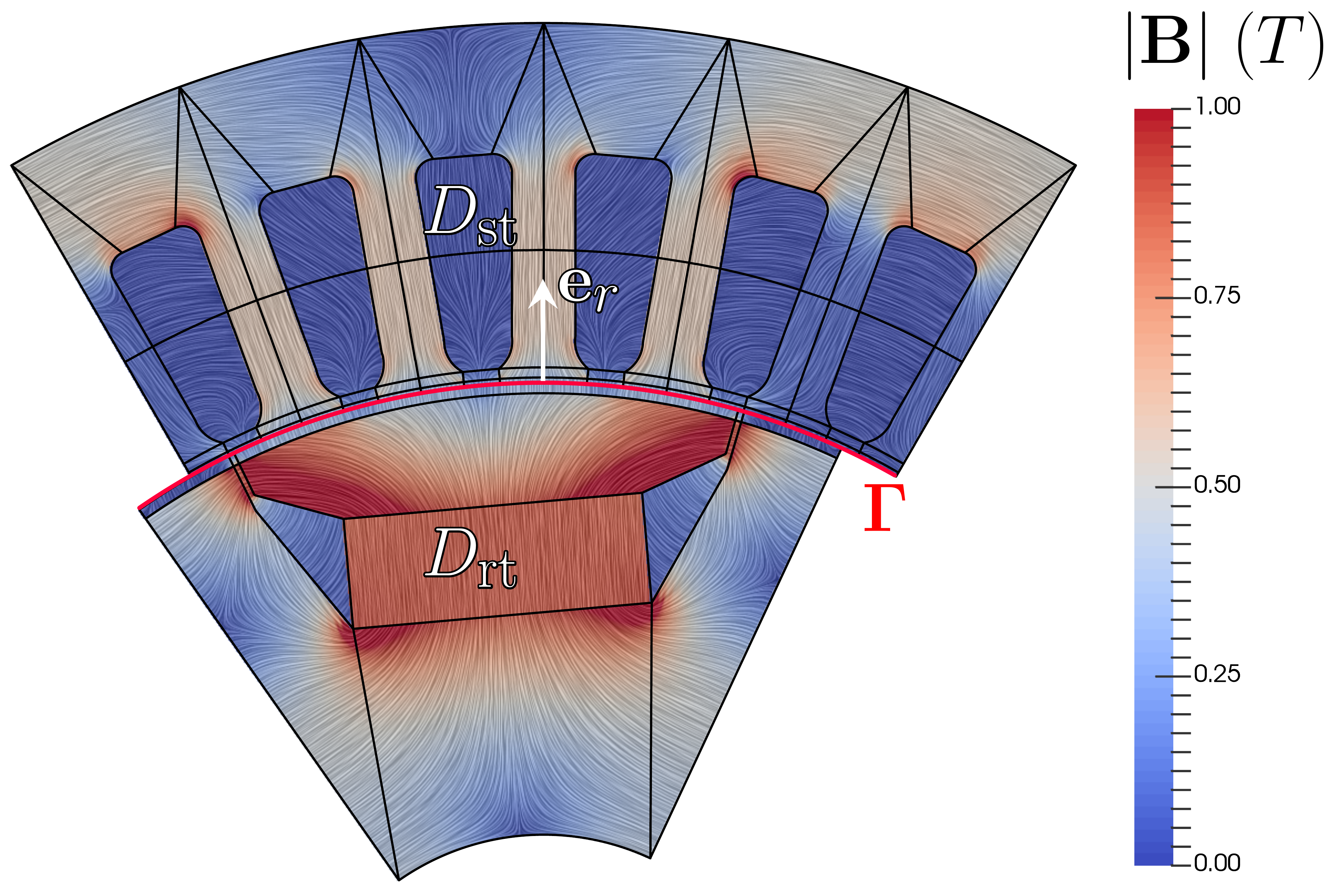}
  \caption{Multipatch model of one pole of a 6-pole permanent magnet synchronous machine. The magnetic flux density $\mathbf{B}$ is shown which was computed using isogeometric analysis.}
  \label{fig:pmsm}
  \vspace{-0.5em}
\end{figure} 
A common quantity of interest when designing electric machines is the total harmonic distortion ($\thd$) of the electromotive force (EMF) ${\mathcal E}$.
 The total harmonic distortion of a function $f=f(t)$ is defined as
 
\begin{align}
\thd_{\setI}(f) = \frac{\sqrt{\sum_{n\in\setI,n\neq1}|c_n|^2}}{|c_1|},
\end{align}
where $c_n$ are the coefficients of the Fourier series of $f(t)$, i.e.
\begin{equation} \label{expFourier1}
    f(t) = \sum_{n=-\infty}^\infty c_n e^{\imath n t},
\end{equation}
and $\setI\subset\setN$ is an index set of frequencies to consider, e.g., one may disregard frequencies that cannot be diminished by shape optimization. Alternatively, the expansion \eqref{expFourier1} can be written as
\begin{align*}
    f(t) = \frac{A_0}{2} + \sum_{n=1}^\infty A_n \cos (nt) + B_n \sin(nt)
\end{align*}
with coefficients $A_n = c_n + c_{-n}$ for $n \geq 0$ and $ B_n = \imath(c_n - c_{-n})$ for $n \geq 1$. Note that it holds that $|c_n| = \sqrt{A_n^2+B_n^2}/2$. Using these coefficients, the THD can be written as
\begin{align} \label{eq_thd_AB}
	\thd(f)=\sqrt{ \frac{\sum_{n\in \setI, n\neq 1} A_n^2+B_n^2}{A_1^2+B_1^2} }.
\end{align}

We are interested in the THD of the EMF. Therefore, the amplitudes $c_n$ are obtained by Fourier analysis 
of the voltages \begin{equation}\label{eq:voltages} 
\mathcal{E}_k(\mathbf{A}(t)) =\partial_t{\Psi}_k(\mathbf{A}(t)),
\end{equation}
with the flux linkage
\begin{equation}\label{eq:fluxlinkage} 
{\Psi}_k(\mathbf{A}(t)) =N_p\int_{D_{\mathrm{3D}}}\pmb{\chi}_k \cdot \mathbf{A}(t)\; {\mathrm{d}} D_{\mathrm{3D}},
\end{equation} 
where $N_p$ is the number of pole pairs. The voltages $\mathcal{E}_k$ are induced by the electromagnetic force in the windings under the assumption that the coils of the poles are connected in series.

In the planar 2D case \eqref{eq:curlcurl} is considered on a cross section ${D} \subset \mathbb R^2$ and reduces to 
\begin{align}
-\nabla\cdot\left(\nu\nabla u\right) = J_{\mathrm{src},z} + \nabla \times \mathbf{M} \cdot  \mathbf{e}_z, \label{eq:reducedmaxwell}
\end{align}
where $u(x,y)$ is the $z$-component  of the magnetic vector potential $\mathbf{A}=[0,0,u]^\top$, $J_{\mathrm{src},z}=\sum_k\chi_ki_k$ is the $z$-component of the current density given by winding functions $\chi_k$ and currents $i_k$ and $\mathbf{e}_z$ is the unit vector in $z$-direction. 

\section{Optimization}\label{sec:opti}

\subsection{Problem statement}
We want to optimize the $2D$ shape $\Omega$ of the 
$3$-phase permanent magnet synchronous machine from Fig.~\ref{fig:pmsm}, which has $N_p=3$ pole pairs, in terms of the total harmonic distortion ($\thd$) of the electromotive force (EMF) as introduced in \eqref{eq:voltages}. Due to periodicity of the stator windings, we will restrict ourselves to the optimization of the first voltage and set $\mathcal E := \mathcal E_1$ and $\chi = \chi_1$. 
Note the dependence of $\mathcal E$ on the shape $\Omega$ via the solution $u$ to \eqref{eq:reducedmaxwell}, i.e., $\mathcal E = \mathcal E(u(t,\Omega))$. Thus, we consider the optimization problem
\begin{equation} 
	\min_{\Omega\in\mathcal{A}} \mathcal{J}(\Omega) := \thd_{\setI}({\mathcal E}(u(t,\Omega))), \label{eq_minTHD}
\end{equation}
that is constrained by the machine model \eqref{eq:reducedmaxwell}, which, for $t \in  [0,T]$, reads in its weak form: Find $u = u(\alpha(t))$ such that
\begin{equation} \label{pde_constraint_weak}
    \underbrace{
	\int_{{D}} \nu_{\Omega}(\alpha(t))\nabla u \cdot\nabla v\,{\mathrm{d}} {D}}
	_{=:a[\alpha(t)](\Omega;u,v)}
	 = 
	 \left\langle r(t, \alpha(t)), v \right\rangle
\end{equation}
for all test functions $v$. Here $T=2\pi/\omega$ is the (electrical) period length, $\alpha(t)$ the rotor angle, which may be given by the equation of motion. We allow all geometries from an admissible set $\mathcal{A}$. Both $u$ and $v$ satisfy homogeneous Dirichlet boundary conditions on the inner and outer circular parts of the boundary and periodic boundary conditions on the left and right parts. Finally, the right-hand-side is given by
\begin{align}
\begin{aligned} 
 \langle r(t, &\alpha(t)), v \rangle = \int_{D_{\mathrm{c}}}\sum_{k=1}^{3}\chi_k(x,y)i_k(t) \, v(x,y) \; {\mathrm{d}} D_{\mathrm{c}} \\
 &+ \int_{D_{\mathrm{pm}}(\alpha(t))} \binom{-M_2}{M_1} \cdot \nabla v(x,y) \; {\mathrm{d}} D_{\mathrm{pm}}(\alpha(t)) \label{eq:rhs}
\end{aligned}
\end{align}
and contains the excitations due to permanent magnets and the coils. Here, $M_1$ and $M_2$ are the first and second component of the magnetization vector $\mathbf{M}$, respectively, $D_{\mathrm{pm}}(\alpha(t))$ denotes the permanent magnet region after rotation by the angle $\alpha(t)$ and we exploit for a more compact notation that the permanent magnets lie in the rotating part and the coils in the stator. We indicate the dependence of the reluctivity function on the current shape $\Omega$ which is subject to the shape optimization by writing $\nu_{\Omega}$.

Moreover, we introduce a temporal discretization $\lbrace t_1, \dots, t_{N_\alpha} \rbrace$ into ${N_{\alpha}}=120$ points in time and a corresponding discretization of the range of angular displacements into ${N_{\alpha}}$ rotor positions, $\boldsymbol \alpha := (\alpha_1,\dots,\alpha_{{N_{\alpha}}})$. For $j \in \{ 1,\dots, {N_{\alpha}} \}$ and a given shape $\Omega$, let $a_j(\Omega;\cdot, \cdot) := a[\alpha_j](\Omega;\cdot,\cdot)$ and $r_j:=r(t_j,\alpha_j)$ according to the definitions in \eqref{pde_constraint_weak} and \eqref{eq:rhs}, respectively. Approximating $u(t,\Omega)$ for $t\in [0,T]$ by $\lbrace u_1, \dots, u_{N_\alpha} \rbrace$ and introducing the notation $\mathcal{J}(u_1, \dots, u_{N_\alpha}) := \thd_{\setI}({\mathcal E}(u_1, \dots, u_{N_\alpha}))$, problem \eqref{eq_minTHD}--\eqref{pde_constraint_weak} after discretization with respect to rotor positions can be written as the optimization problem
\begin{equation}
	\qquad\underset{\Omega}{\mbox{min }} \mathcal{J}(u_1, \dots, u_{{N_{\alpha}}}) \label{minJ}
\end{equation} \vspace{-6mm}
\begin{align} \label{pdeconstraints}
	\mbox{s.t. } \left\lbrace \begin{array}{c}
					\;\quad a_1(\Omega;u_1,v_1) = \langle r_1, v_1 \rangle \; \forall v_1,\\
					\;\;\qquad\vdots\\
					a_{{N_{\alpha}}}(\Omega;u_{{N_{\alpha}}},v_{{N_{\alpha}}}) = \langle r_{{N_{\alpha}}}, v_{{N_{\alpha}}} \rangle  \; \forall  v_{{N_{\alpha}}},
	                          \end{array} \right. 
\end{align}
which is constrained by ${N_{\alpha}}$ boundary value problems corresponding to the ${N_{\alpha}}$ rotor positions under consideration.

\subsection{Shape Sensitivity Analysis}
The shape derivative $d \mathcal{J}(\Omega; \mathbf{W})$ of a domain-dependent functional $\mathcal{J}= \mathcal{J}(\Omega)$ represents the sensitivity of the functional with respect to a perturbation of the domain in the direction of a given vector field $\mathbf{W}$. The shape derivative is defined as
\begin{equation}
	d \mathcal{J}(\Omega; \mathbf{W}) = \underset{\delta \searrow 0}{\mbox{lim}} \frac{\mathcal{J}(T_\delta^{\mathbf{W}}(\Omega)) - \mathcal{J}(\Omega)}{\delta},
\end{equation}
if this limit exists and the mapping $\mathbf{W} \mapsto d \mathcal{J}(\Omega; \mathbf{W})$ is linear and continuous on the space of smooth vector fields. Here, $T_\delta^{\mathbf{W}}$ represents a transformation which moves every point $\x$ a distance $\delta>0$ in the direction given by the vector field $\mathbf{W}$, $T_\delta^{\mathbf{W}}(\x) = \x + \delta \mathbf{W}(\x)$.

For deriving the shape derivative $d\mathcal{J}(\Omega; \mathbf{W})$ for the optimization problem \eqref{minJ}--\eqref{pdeconstraints}, we follow the steps taken in \cite{GanglLangerLaurainMeftahiSturm2015}. First, we introduce the Lagrangian
\begin{align*}
	\mathcal L(\Omega, &\varphi_1,\dots,\varphi_{{N_{\alpha}}},\psi_1,\dots, \psi_{{N_{\alpha}}}) := \\
	& \mathcal{J}(\varphi_1, \dots, \varphi_{{N_{\alpha}}}) + \sum_{k=1}^{{N_{\alpha}}} \left(a_k(\varphi_k,\psi_k) - \langle r_k, \psi_k \rangle \right).
\end{align*}
For ease of notation, we will use the notation $\mathbf{u} = (u_1,\dots, u_{{N_{\alpha}}})$ and similar for other quantities.
Note that, for $u_i$ satisfying the $i$-th equation of \eqref{pdeconstraints}, it holds that
\begin{align*}
	\frac{\partial}{\partial \psi_i} \mathcal L(\Omega, \mathbf{u},\boldsymbol \psi)(q_i) 
	 = a_i(u_i,q_i) - \langle r_i,q_i\rangle = 0
\end{align*}
for any test function $q_i$. Similarly, we introduce the adjoint states $p_i$, $i=1,\dots,{N_{\alpha}}$, as the solutions to 
\begin{align} \label{eq_adjoint}
	0= \frac{\partial}{\partial \varphi_i} \mathcal L(\Omega, \mathbf{u},\mathbf{p})(v_i) =  \frac{\partial \mathcal{J}}{\partial \varphi_i}(\mathbf{u})(v_i) + a_i(v_i,p_i)
\end{align}
for all test functions $v_i$.

Let us now consider the functional $\mathcal{J}= \mathcal{J}(u_1, \dots, u_{{N_{\alpha}}})$ more closely. We will use the representation \eqref{eq_thd_AB} for the THD.
{Let $(C_n)_{n \in \setI}$ the Fourier coefficients of the function ${\Psi}(u_1, \dots, u_{{N_{\alpha}}})$, i.e.,
\begin{equation}\label{eq:flux_fourier}
	{\Psi}(u_1, \dots, u_{{N_{\alpha}}}) =  \sum_{n=-{N_{\alpha}}}^{N_{\alpha}} C_n e^{\imath n t}.
\end{equation}
Using the Fourier representation \eqref{eq:flux_fourier} of the flux linkage and \eqref{eq:voltages}, the electromotive force can be written as 
\begin{equation}
	{\mathcal E}(u_1, \dots, u_{{N_{\alpha}}}) =  \sum_{n=-{N_{\alpha}}}^{N_{\alpha}} \underbrace{\imath n C_n}_{c_n} e^{\imath n t},
\end{equation}
where $(c_n)_{n \in \setI}$ are the Fourier coefficients of the electromotive force ${\mathcal E}$.
This Fourier series can be rewritten as
\begin{equation}
	{\mathcal E}(u_1, \dots, u_{{N_{\alpha}}}) = \frac{A_0}{2} + \sum_{n=1}^{N_{\alpha}} A_n \cos (nt) + B_n \sin(nt). \label{eq:fourierSeries}
\end{equation}
The coefficients $A_n$ and $B_n$ of the Fourier representation of $\partial_{t}{\Psi}$ \eqref{eq:fourierSeries} are obtained by time derivation and Fourier transform, which we will denote by $A_n=[\mathcal{F'}_a({\Psi}(u_1, \dots, u_{{N_{\alpha}}}))]_n$ and $B_n=[\mathcal{F'}_b({\Psi}(u_1, \dots, u_{{N_{\alpha}}}))]_n$, respectively. Exploiting the linearity 
of the discrete Fourier transform, the vectors $\mathbf{A}$ and $\mathbf{B}$ can also be written in terms of transformation matrices $\mathbf{M}_a = \mathcal{F'}_a({\mathbf{I}})$ and $\mathbf{M}_b=\mathcal{F'}_b({\mathbf{I}})$, where ${\mathbf{I}}$ is the identity matrix, i.e., 
\begin{align}
	A_k(u_1,\dots,u_{{N_{\alpha}}}) &= \sum_{j=1}^{N_{\alpha}}\left( \left(\mathbf{M}_a\right)_{k,j} N_p l_z\int_{{D}}\chi u_j\; {\mathrm{d}} {D} \right),\label{eq:def_Ak}\\
	B_k(u_1,\dots,u_{{N_{\alpha}}}) &= \sum_{j=1}^{N_{\alpha}} \left( \left(\mathbf{M}_a\right)_{k,j} N_p l_z\int_{{D}}\chi u_j\; {\mathrm{d}} {D} \right),\label{eq:def_Bk}
\end{align}
where $l_z$ is the length of the machine in $z$-direction.
}

In order to solve the adjoint state equation \eqref{eq_adjoint}, we need to differentiate the functional $\mathcal{J} = \thd({\mathcal E}(u_1,\dots,u_{{N_{\alpha}}}))$ with respect to $u_i$ for $i \in \{1,\dots,{N_{\alpha}}\}$. Using the relation \eqref{eq_thd_AB}, we get
\begin{align*}
	&\frac{d\, \thd({\mathcal E}(\mathbf{u}))}{d u_i}(\mathbf{u})(v_i) = \\
	&\frac{1}{\sqrt{A_1^2+B_1^2}}  \frac{1}{ \sqrt{\sum_{k\in \setI, k\neq 1} A_k^2 + B_k^2} } \left(\sum_{k\in \setI, k\neq 1} A_k A_k' + B_k B_k'\right) \\
	&- \frac{1}{(A_1^2+B_1^2)^{3/2}} \left( \sqrt{\sum_{k\in \setI, k\neq 1} A_k^2+B_k^2} \left( A_1 A_1' + B_1 B_1'\right) \right).
\end{align*}
Here, we used the abbreviations $A_k' := \frac{d A_k}{d u_i}(\mathbf{u})(v_i)$ and $B_k' := \frac{d B_k}{d u_i}(\mathbf{u})(v_i)$.
It can be seen from \eqref{eq:def_Ak} and \eqref{eq:def_Bk} that
\begin{align*}
    \frac{d A_k}{d u_i}(\mathbf{u})(v_i) &= \left(\mathbf{M}_a\right)_{k,i} N_p l_z \int_D \chi v_i \; {\mathrm{d}} {D}, \\
    \frac{d B_k}{d u_i}(\mathbf{u})(v_i) &= \left(\mathbf{M}_b\right)_{k,i} N_p l_z \int_D \chi v_i \; {\mathrm{d}} {D}.
\end{align*}
Given the solutions to the forward problem \eqref{pde_constraint_weak} for all rotor positions $l \in \{1,\dots {N_{\alpha}}\}$, we obtain for the adjoint problem \eqref{eq_adjoint} defining the adjoint variable at rotor position $l \in \{1, \dots {N_{\alpha}}\}$: Find $p_i$ such that
\begin{align*}
	a_i(v_i, p_i) = - \frac{d\, \thd({\mathcal E}(\mathbf{u}))}{d u_i}(\mathbf{u})(v_i) 
\end{align*}
for all test functions $v_i$.

Finally, in a similar way as it was proposed in \cite{GanglLangerLaurainMeftahiSturm2015}, assuming that the deformation vector field $\mathbf{W}$ vanishes on the interface $\Gamma$ and on the stator, and that the permanent magnet region remains unchanged, we obtain the formula for the shape derivative in the direction of a smooth vector field $\mathbf{W} \in C^1(D,\mathbb R^2)$:
\begin{align} \label{eq_shape_der}
	\begin{aligned}
	 	d\mathcal{J}&(\Omega; \mathbf{W}) = \sum_{l=1}^{N_{\alpha}} \int_{\Omega_{\mathrm{pm}}} (\nabla \cdot (\mathbf{W}) \mathbf{I} - \mathbf{\mathbf{DW}}^\top) \nabla p_l \cdot \binom{-M_2}{M_1} \;  {\mathrm{d}} x \\
		&+ \sum_{l=1}^{N_{\alpha}}\int_D \nu \, (\nabla \cdot \mathbf{W} \, \mathbf{I} - \mathbf{\mathbf{DW}}^\top - \mathbf{\mathbf{DW}}) \nabla u_l \cdot \nabla p_l \; {\mathrm{d}} x.		
	\end{aligned}
\end{align}
Here, $\mathbf{I} \in \mathbb R^{2\times 2}$ denotes the two-dimensional identity matrix.

\section{Solution of the Forward Problem by IGA} \label{sec:discretization}
In this section, we will treat the numerical solution of the forward problem \eqref{eq:reducedmaxwell} using Isogeometric Analysis and harmonic stator-rotor coupling allowing for a flexible treatment of the rotation.
\subsection{Harmonic Stator-Rotor Coupling}
As proposed in \cite{Bontinck_2018ac}, stator and rotor domains ${D}_q$, where $q\in \{ \text{st},\text{rt} \}$ distinguishes between stator and rotor and $\overline{ {D}}=  \overline {D}_{\mathrm{rt}} \cup \overline  {D}_{\mathrm{st}}$, are considered separately. They are coupled at the stator-rotor interface $\Gamma={\overline{{D}}_{\mathrm{rt}}} \cap {\overline{{D}}_{\mathrm{st}}}$ in the air gap by enforcing the continuity of the magnetic vector potential $u$ and of the azimuthal component of the magnetic field strength $H_{\theta}^{(\mathrm{st})}|_{\Gamma}(\theta_{\mathrm{st}})=H_{\theta}^{(\mathrm{rt})}|_{\Gamma}(\theta_{\mathrm{rt}})$, where $\theta_{\mathrm{st}}$ and $\theta_{\mathrm{rt}}$ are the angular coordinates attached to stator and rotor domain, respectively. The angular displacement between the domains is $\alpha=\theta_{\mathrm{st}}-\theta_{\mathrm{rt}}$.
In its weak form the problem \eqref{eq:reducedmaxwell} can be formulated as:

Find $({{u}_{\mathrm{st}}},{{u}_{\mathrm{rt}}},\lambda) \in {V_{\mathrm{st}}} \times {V_{\mathrm{rt}}} \times{\Lambda}$ such that
\begin{align*}
 \sum_q \int\limits_{{D}_{q}} \nu_{q}{\nabla u_{q}} \cdot {\nabla v_{q}} \operatorname{d} {D}_{q} + \int\limits_{\Gamma} \lambda \left\llbracket v\right\rrbracket \operatorname{d}\Gamma&= {\sum_{q} \langle r_q, v_q \rangle}\\
\intertext{with interface condition} \int_\Gamma \left\llbracket u\right\rrbracket \mu \operatorname{d}\Gamma &= 0, 
\end{align*}
for all $v_q \in V_q$ and $\mu \in \Lambda$, $q\in \{ \mathrm{st},\mathrm{rt} \}$, where $V_q$ are suitable function spaces which contain the necessary Dirichlet boundary conditions, $v_q$ are the test functions, 
and $r_q$, $q \in \{ \mathrm{st}, \mathrm{rt} \}$ are the contributions of the right hand side on the stator and rotor as in \eqref{eq:rhs},
\begin{align}
    \langle r_{\mathrm{st}}, v_{\mathrm{st}} \rangle &= \int_{D_{\mathrm{c}}} \sum_k\chi_{k}(x,y)i_k(t) \, v_{\mathrm{st}}(x,y) \;{\mathrm{d}} D_{\mathrm{c}}, \\
    \langle r_{\mathrm{rt}}, v_{\mathrm{rt}} \rangle &= \int_{D_{\text{pm}}(\alpha)}  \binom{-M_2}{M_1} \cdot \nabla v(x,y) \;{\mathrm{d}} D_{\text{pm}}(\alpha) .
\end{align}

Moreover, $\llbracket v\rrbracket =(v_{\textrm{st}}-v_{\textrm{rt}})\big |_{\Gamma}$ denotes the jump of $v$ across the interface $\Gamma$ and $\lambda = \nu{\nabla u^{(\mathrm{st})}}\cdot{\bm{\mathbf{e}}}_{r} = \nu{\nabla u^{(\mathrm{rt})}}\cdot{\bm{\mathbf{e}}}_{r} = H_\theta^{(q)}$, where ${\bm{\mathbf{e}}}_{r}$ is the unit vector in radial direction, is the $\theta$-component of the magnetic field at the interface and can be interpreted as a Lagrange multiplier \cite{Hansbo_2005aa}.

\subsection{Discretization}
 The $z$-component of the magnetic vector potential $u$ is discretized by a linear combination of the same scalar basis functions $w^{(q)}_{j}$ used for weighting and $H_{\theta}$ is expressed by a superposition of ${N_\Gamma}$ harmonic basis functions \cite{De-Gersem_2004ad} as
\begin{align}\label{eq:basis}
 u^{(q)}\approx \sum \limits_{j=1}^{N_q} u^{(q)}_{j} w^{(q)}_{j}, \quad
 H_{\theta}^{(q)}(\theta_q) \approx \sum_{k=1}^{{N_\Gamma}} \lambda_{k}^{(q)} e^{-\imath\ell_k\theta_q},
\end{align}
where $\bm{\lambda}^{(q)}\in \mathbb{R}^{{N_\Gamma}}$ is the vector of Fourier coefficients for the harmonic basis functions, $\ell_k$ are the harmonic orders and $\mathbf{u}^{(q)}\in\mathbb{R}^{N_{q}}$ is the vector of degrees of freedom for each domain. This allows for a more efficient simulation of angular displacements $\alpha=\theta_{\mathrm{st}}-\theta_{\mathrm{rt}}$ when compared to classical mortaring, moving-band or sliding-surface methods~\cite{De-Gersem_2004af}. Note that $u(\alpha)=u(x,y;\alpha)$  with $(x,y)\in{D}$, and the degrees of freedom $\mathbf{u}^{(q)}(\alpha)$ depend on the rotor position. 
The approach of harmonic stator-rotor coupling in an IGA framework leads to the Mortar-type saddle-point problem \cite{Bontinck_2018ac},
\begin{equation}
\begin{bmatrix}
	    {{\mathbf{K}}_{\mathrm{st}}}	&	0	&	-{{\mathbf{G}}_{\mathrm{st}}}\mathbf{R}(\alpha)\\
		0	&	{{\mathbf{K}}_{\mathrm{rt}}}	&	{{\mathbf{G}}_{\mathrm{rt}}}\\
-\mathbf{R}(\alpha){{\mathbf{G}}_{\mathrm{st}}^H}	&	{{\mathbf{G}}_{\mathrm{rt}}^H}	&	0 
\end{bmatrix}
\hspace{-0.1cm}
\begin{bmatrix}
{{\mathbf{u}}^{(\mathrm{st})}}\\
{{\mathbf{u}}^{(\mathrm{rt})}}\\
\bm{\lambda}
\end{bmatrix}=
\begin{bmatrix}
{{\mathbf{j}}_{\mathrm{st}}}\\
{{\mathbf{j}}_{\mathrm{rt}}}\\
0
\end{bmatrix}. \label{eq:system}
\end{equation}
Here, $\bm{\lambda}=\bm{\lambda}^{(\mathrm{st})}=\bm{\lambda}^{(\mathrm{rt})}$ is enforced strongly by testing with $v=w$, 
$\mathbf{K}_q \in \mathbb{R}^{N_{q} \times N_{q}}$ are the stiffness matrices with
\begin{equation}
 (\mathbf K_q)_{ij} = \int_{{D}_{q}} \left(\nu{\frac{\partial w_i}{\partial x}}{\frac{\partial w_j}{\partial x}}+\nu{\frac{\partial w_i}{\partial y}}{\frac{\partial w_j}{\partial y}} \right)\;\text{d}{D}_{q},
\end{equation}
$\mathbf{G}_q \in \mathbb{C}^{N_{q} \times {N_\Gamma}}$ are the coupling matrices with
\begin{align}
(\mathbf{G}_{\mathrm{st}})_{ik} &=-\int_{0}^{2\pi}e^{-\imath\ell_k\theta_{\mathrm{st}}} w_i\left(\mathbf{r}\left(\theta_{\mathrm{st}}\right)\right)R_{\Gamma}\;{\mathrm{d}}\theta_{\mathrm{st}}, \\
(\mathbf{G}_{\mathrm{rt}})_{ik} &= \int_{0}^{2\pi}e^{-\imath\ell_k\theta_{\mathrm{rt}}}w_i\left(\mathbf{r}\left(\theta_{\mathrm{rt}}\right)\right)R_{\Gamma}\;{\mathrm{d}}\theta_{\mathrm{rt}},
\end{align}
where
$R_{\Gamma}$ is the radius of the interface $\Gamma$, $\mathbf{r}(\theta)=\left(R_{\Gamma}\cos(\theta), R_{\Gamma}\sin(\theta)\right)^{\top}$ is a mapping from an angle $\theta$ to the point on the interface at this angle in Cartesian coordinates and 
$\mathbf{R}(\alpha) \in \mathbb{C}^{{N_\Gamma} \times {N_\Gamma}}$ is the diagonal rotation matrix with
$\left(\mathbf{R}(\alpha)\right)_{kk}=e^{\imath\ell_k\alpha}$. The saddle-point problem \eqref{eq:system} is stable if the numbers $N_q$ and ${N_\Gamma}$ of basis functions are chosen consistently \cite{Bontinck_2018ac}.  
\subsection{Elimination of Inner Degrees of Freedom}
The system \eqref{eq:system} can be rewritten as
\begin{align}
\begin{bmatrix}
\mathbf{I} & 0 & -{{\mathbf{K}}_{\mathrm{st}}}^{-1}{{\mathbf{G}}_{\mathrm{st}}}\mathbf{R}(\alpha)\\
0 & \mathbf{I} & {{\mathbf{K}}_{\mathrm{rt}}}^{-1}{{\mathbf{G}}_{\mathrm{rt}}}\\
-\mathbf{R}(\alpha){{\mathbf{G}}_{\mathrm{st}}^H} & {{\mathbf{G}}_{\mathrm{rt}}}^H & 0\\
\end{bmatrix}
\begin{bmatrix}
{{\mathbf{u}}^{(\mathrm{st})}}\\
{{\mathbf{u}}^{(\mathrm{rt})}}\\
\bm{\lambda}
\end{bmatrix}
&=
\begin{bmatrix}
{{\mathbf{K}}_{\mathrm{st}}}^{-1}{{\mathbf{j}}_{\mathrm{st}}}\\{{\mathbf{K}}_{\mathrm{rt}}}^{-1}{{\mathbf{j}}_{\mathrm{rt}}}\\0
\end{bmatrix}.
\end{align}
The internal degrees of freedom can be eliminated using the \emph{Schur-complement}, giving rise to the low-dimensional \emph{interface problem}
\begin{align}
\mathbf{K}_{\textrm{int}}(\alpha)
\bm{\lambda}^{(\mathrm{rt})}
&=
\mathbf{f}_{\textrm{int}}(\alpha) \label{eq:sysint}
\end{align}
with
\begin{align}
\mathbf{K}_{\textrm{int}}(\alpha)&={{\mathbf{G}}_{\mathrm{rt}}^H} {{{\mathbf{K}}_{\mathrm{rt}}}^{-1}{{\mathbf{G}}_{\mathrm{rt}}}} + \mathbf{R}(\alpha){{\mathbf{G}}_{\mathrm{st}}^H} {{{\mathbf{K}}_{\mathrm{st}}}^{-1}{{\mathbf{G}}_{\mathrm{st}}}}\mathbf{R}(\alpha), \label{eq:kint}
	\\
	\mathbf{f}_{\textrm{int}}(\alpha)&={{\mathbf{G}}_{\mathrm{rt}}}^H {{{\mathbf{K}}_{\mathrm{rt}}}^{-1}{{\mathbf{j}}_{\mathrm{rt}}}} - \mathbf{R}(\alpha){{\mathbf{G}}_{\mathrm{st}}^H} {{{\mathbf{K}}_{\mathrm{st}}}^{-1}{{\mathbf{j}}_{\mathrm{st}}}}. \label{eq:fint} 
\end{align}
The inverses in \eqref{eq:kint} and \eqref{eq:fint} are not needed explicitly. Instead, one factorization and a few forward/backward substitutions (for each spectral basis at the interface) can be used to precompute the necessary expressions, e.g., $\mathbf{K}_{*}^{-1}\mathbf{G}_{*}$, which are independent of $\alpha$. 
Thus, only the small system \eqref{eq:sysint} with the system matrix $\mathbf{K}_{\textrm{int}} \in \mathbb{C}^{{N_\Gamma} \times {N_\Gamma}}$ has to be solved in the online phase for different rotation angles. 

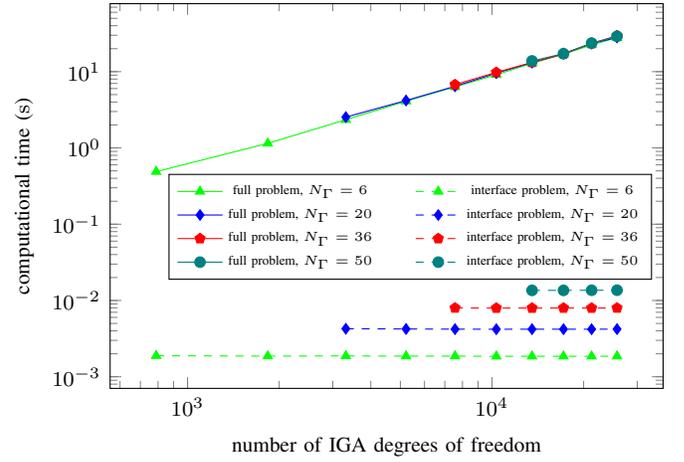
\begin{figure}
\centering
\begin{tikzpicture}[]
           \begin{loglogaxis}[width=\columnwidth, height=0.75\columnwidth, ylabel={computational time (\si{s})},xlabel={number of IGA degrees of freedom}, ylabel near ticks,tick label style={font=\footnotesize}, legend columns=4,transpose legend, legend style={font=\tiny, /tikz/every even column/.append style={column sep=0.5cm}}, label style={font=\footnotesize},every x tick scale label/.style={at={(1,0)},anchor=north,yshift=-5pt,inner sep=0pt},legend style={at={(0.98,0.42)},anchor=east}]
                \addplot [green, mark=triangle*, mark options={solid}] table [x index=0, y index=1, col sep=comma] {images/dofs_timeFullNharm_6_20_36_50_std.csv};
                \addplot [blue, mark=diamond*, mark options={solid}] table [x index=0, y index=2, col sep=comma] {images/dofs_timeFullNharm_6_20_36_50_std.csv};
                \addplot [red, mark=pentagon*, mark options={solid}] table [x index=0, y index=3, col sep=comma] {images/dofs_timeFullNharm_6_20_36_50_std.csv};
                \addplot [teal, mark=*, mark options={solid}] table [x index=0, y index=4, col sep=comma] {images/dofs_timeFullNharm_6_20_36_50_std.csv};
                \addplot [green, mark=triangle*, mark options={solid}, dashed] table [x index=0, y index=1, col sep=comma] {images/dofs_timeSchurNharm_6_20_36_50_std.csv};
                \addplot [blue, mark=diamond*, mark options={solid}, dashed] table [x index=0, y index=2, col sep=comma] {images/dofs_timeSchurNharm_6_20_36_50_std.csv};
                \addplot [red, mark=pentagon*, mark options={solid}, dashed] table [x index=0, y index=3, col sep=comma] {images/dofs_timeSchurNharm_6_20_36_50_std.csv};
                \addplot [teal, mark=*, mark options={solid}, dashed] table [x index=0, y index=4, col sep=comma] {images/dofs_timeSchurNharm_6_20_36_50_std.csv};
                \legend{full problem, ${N_\Gamma}=6$\\full problem, ${N_\Gamma}=20$\\full problem, ${N_\Gamma}=36$\\full problem, ${N_\Gamma}=50$\\interface problem, ${N_\Gamma}=6$\\interface problem, ${N_\Gamma}=20$\\interface problem, ${N_\Gamma}=36$\\interface problem, ${N_\Gamma}=50$\\}
            \end{loglogaxis}
\end{tikzpicture} \caption{Comparison of the computational time of the simulation of a rotating machine by solving the interface problem \cref{eq:sysint} compared to the computational time of the simulation solving the full system \cref{eq:system}.}
 \label{fig:schur_vs_full}
 \end{figure}

The internal degrees of freedom can be cheaply reconstructed, i.e.,
\begin{align}
	{{\mathbf{u}}^{(\mathrm{st})}}&={{{\mathbf{K}}_{\mathrm{st}}}^{-1}{{\mathbf{j}}_{\mathrm{st}}}}+{{{\mathbf{K}}_{\mathrm{st}}}^{-1}{{\mathbf{G}}_{\mathrm{st}}}}\mathbf{R}(\alpha)\bm{\lambda}^{(\mathrm{rt})},\label{eq:recostructionust}\\
	{{\mathbf{u}}^{(\mathrm{rt})}}&={{{\mathbf{K}}_{\mathrm{rt}}}^{-1}{{\mathbf{j}}_{\mathrm{rt}}}}-{{{\mathbf{K}}_{\mathrm{rt}}}^{-1}{{\mathbf{G}}_{\mathrm{rt}}}}\bm{\lambda}^{(\mathrm{rt})}.\label{eq:recostructionurt}	
\end{align}
This is computationally convenient when dealing with rotation since only the low-dimensional matrix $\mathbf{R}(\alpha)$ depends on $\alpha$. This leads to a significant reduction in computational cost of the shape optimization, e.g., during the computation of the total harmonic distortion, where \eqref{eq:system} has to be solved for $N_{\alpha}$ rotor positions. The computational cost of the simulation of a rotor rotation of $\SI{120}{\degree}$ with $N_{\alpha}=120$ discretized points in time using the interface problem is compared to the computational cost using the full system in \cref{fig:schur_vs_full}. It can be seen that the computational time needed to solve the full system \cref{eq:system} depends on the number of the IGA degrees of freedom whereas the computational time needed to solve the interface problem \cref{eq:sysint} is almost independent of the number of IGA degrees of freedom and depends on the number of harmonics that are used for the coupling.  The rotor is rotated $\SI{120}{\degree}$ using $N_{\alpha}=120$ time steps. The simulation is carried out for different spatial refinements and for different numbers of harmonics ${N_\Gamma}$ for the coupling. The preprocessing, i.e., computing the inverses in \cref{eq:kint,eq:fint}, took at most \SI{0.678}{s} in the case of ${N_\Gamma}=50$ and $25746$ IGA DoFs and the postprocessing, i.e., \cref{eq:recostructionust,eq:recostructionurt}, took at most $\SI{1.3012}{s}$. The computation is is repeated 100 times and the computational time is averaged. The computation is carried out in Matlab\textsuperscript{\textregistered} R2019a on a 6-core machine (Intel\textsuperscript{\textregistered}
 Core\texttrademark{} i7-5820K CPU) with \SI{16}{GB} RAM.

\subsection{B-splines and NURBS} \label{sec:splines}
\begin{figure}
 \center
 \begin{subfigure}[c]{0.49\columnwidth}
  \resizebox{\textwidth}{!}{\definecolor{mycolor1}{rgb}{0.00000,0.44700,0.74100}\begin{tikzpicture}

\begin{axis}[width=4.519in,
height=3.564in,
at={(0.758in,0.481in)},
scale only axis,
axis lines=none,
xmin=-0.1,
xmax=3.1,
ymin=-0.1,
ymax=2.1,
axis background/.style={fill=white}
]
\addplot [color=mycolor1, line width=2.0pt, forget plot]
  table[row sep=crcr]{0	0\\
0.000935049480375794	0.073501498215216\\
0.00369261270621687	0.143326245462644\\
0.0082013118545938	0.209585273911285\\
0.0143897691025771	0.272389615730141\\
0.0221866066272373	0.331850303088213\\
0.0315204466056451	0.388078368154502\\
0.0423199112148708	0.44118484309801\\
0.0545136226319852	0.491280760087738\\
0.0680302030340587	0.538477151292687\\
0.0827982745981619	0.58288504888186\\
0.0987464595013653	0.624615485024256\\
0.115803379920739	0.663779491888878\\
0.133897658033355	0.700488101644727\\
0.152957916016283	0.734852346460804\\
0.172912776046592	0.766983258506111\\
0.193690860301355	0.796991869949649\\
0.215220790957642	0.824989212960419\\
0.237431190192522	0.851086319707423\\
0.260250680183067	0.875394222359662\\
0.283607883106348	0.898023953086138\\
0.307431421139434	0.919086544055852\\
0.331649916459396	0.938693027437804\\
0.356191991243305	0.956954435400998\\
0.380986267668232	0.973981800114433\\
0.405961367911246	0.989886153747112\\
0.431045914149419	1.00477852846803\\
0.456168528559821	1.0187699564462\\
0.481257833319523	1.03197146985062\\
0.506242450605595	1.04449410085029\\
0.531051002595107	1.0564488816142\\
0.555612111465131	1.06794684431137\\
0.579854399392736	1.07909902111079\\
0.603706488554994	1.09001644418146\\
0.627097001128975	1.10081014569239\\
0.649954559291749	1.11159115781258\\
0.672207785220387	1.12247051271103\\
0.69378530109196	1.13355924255673\\
0.714615729083538	1.1449683795187\\
0.734627691372191	1.15680895576593\\
0.753749958177883	1.16919181312656\\
0.77194268481368	1.18218744116617\\
0.789236050880565	1.19577629822302\\
0.805669710724612	1.20992666082024\\
0.821283318691891	1.22460680548099\\
0.836116529128475	1.2397850087284\\
0.850208996380436	1.25542954708561\\
0.863600374793845	1.27150869707578\\
0.876330318714775	1.28799073522203\\
0.888438482489297	1.30484393804752\\
0.899964520463482	1.32203658207538\\
0.910948086983405	1.33953694382876\\
0.921428836395135	1.3573132998308\\
0.931446423044746	1.37533392660465\\
0.941040501278308	1.39356710067344\\
0.950250725441894	1.41198109856032\\
0.959116749881576	1.43054419678842\\
0.967678228943426	1.44922467188091\\
0.975974816973515	1.4679908003609\\
0.984046168317916	1.48681085875156\\
0.9919319373227	1.50565312357601\\
0.99967177833394	1.52448587135741\\
1.00730534569771	1.54327737861889\\
1.01487229376007	1.5619959218836\\
1.02241227686711	1.58060977767468\\
1.02996494936489	1.59908722251528\\
1.03756996559948	1.61739653292852\\
1.04526697991697	1.63550598543757\\
1.05309564666341	1.65338385656556\\
1.06109562018488	1.67099842283562\\
1.06930655482745	1.68831796077092\\
1.0777681049372	1.70531074689458\\
1.0865199248602	1.72194505772975\\
1.09560166894251	1.73818916979957\\
1.10505299153021	1.75401135962719\\
1.11491354696938	1.76937990373574\\
1.12522298960608	1.78426307864838\\
1.13602097378638	1.79862916088824\\
1.14734715385637	1.81244642697846\\
1.1592411841621	1.82568315344219\\
1.17174195768621	1.83830795518632\\
1.18485876940742	1.85030260178615\\
1.19856246545049	1.86166595119652\\
1.21282132233857	1.87239800341742\\
1.2276036165948	1.88249875844886\\
1.24287762474232	1.89196821629083\\
1.25861162330427	1.90080637694334\\
1.2747738888038	1.90901324040639\\
1.29133269776405	1.91658880667997\\
1.30825632670816	1.92353307576408\\
1.32551305215928	1.92984604765873\\
1.34307115064054	1.93552772236391\\
1.36089889867509	1.94057809987963\\
1.37896457278607	1.94499718020589\\
1.39723644949663	1.94878496334268\\
1.4156828053299	1.95194144929\\
1.43427191680903	1.95446663804786\\
1.45297206045716	1.95636052961626\\
1.47175151279744	1.95762312399519\\
1.490578550353	1.95825442118465\\
1.509421449647	1.95825442118465\\
1.52824848720256	1.95762312399519\\
1.54702793954284	1.95636052961626\\
1.56572808319097	1.95446663804786\\
1.5843171946701	1.95194144929\\
1.60276355050337	1.94878496334268\\
1.62103542721393	1.94499718020589\\
1.63910110132491	1.94057809987963\\
1.65692884935946	1.93552772236391\\
1.67448694784072	1.92984604765873\\
1.69174367329184	1.92353307576408\\
1.70866730223595	1.91658880667997\\
1.7252261111962	1.90901324040639\\
1.74138837669573	1.90080637694334\\
1.75712237525768	1.89196821629083\\
1.7723963834052	1.88249875844886\\
1.78717867766143	1.87239800341742\\
1.80143753454951	1.86166595119652\\
1.81514123059258	1.85030260178615\\
1.82825804231379	1.83830795518632\\
1.8407588158379	1.82568315344219\\
1.85265284614363	1.81244642697846\\
1.86397902621362	1.79862916088824\\
1.87477701039392	1.78426307864838\\
1.88508645303062	1.76937990373574\\
1.89494700846979	1.75401135962719\\
1.90439833105749	1.73818916979957\\
1.9134800751398	1.72194505772975\\
1.9222318950628	1.70531074689458\\
1.93069344517255	1.68831796077092\\
1.93890437981512	1.67099842283562\\
1.94690435333659	1.65338385656556\\
1.95473302008303	1.63550598543757\\
1.96243003440052	1.61739653292853\\
1.97003505063511	1.59908722251528\\
1.97758772313289	1.58060977767468\\
1.98512770623993	1.5619959218836\\
1.99269465430229	1.54327737861889\\
2.00032822166606	1.52448587135741\\
2.0080680626773	1.50565312357601\\
2.01595383168208	1.48681085875156\\
2.02402518302648	1.4679908003609\\
2.03232177105657	1.44922467188091\\
2.04088325011842	1.43054419678843\\
2.04974927455811	1.41198109856032\\
2.05895949872169	1.39356710067344\\
2.06855357695525	1.37533392660465\\
2.07857116360486	1.3573132998308\\
2.0890519130166	1.33953694382876\\
2.10003547953652	1.32203658207538\\
2.1115615175107	1.30484393804752\\
2.12366968128523	1.28799073522203\\
2.13639962520615	1.27150869707578\\
2.14979100361956	1.25542954708561\\
2.16388347087152	1.2397850087284\\
2.17871668130811	1.22460680548099\\
2.19433028927539	1.20992666082024\\
2.21076394911943	1.19577629822302\\
2.22805731518632	1.18218744116617\\
2.24625004182212	1.16919181312656\\
2.26537230862781	1.15680895576593\\
2.28538427091646	1.1449683795187\\
2.30621469890804	1.13355924255673\\
2.32779221477961	1.12247051271103\\
2.35004544070825	1.11159115781258\\
2.37290299887103	1.10081014569239\\
2.39629351144501	1.09001644418146\\
2.42014560060726	1.07909902111079\\
2.44438788853487	1.06794684431137\\
2.46894899740489	1.0564488816142\\
2.4937575493944	1.04449410085029\\
2.51874216668048	1.03197146985062\\
2.54383147144018	1.0187699564462\\
2.56895408585058	1.00477852846804\\
2.59403863208875	0.989886153747112\\
2.61901373233177	0.973981800114433\\
2.64380800875669	0.956954435400998\\
2.6683500835406	0.938693027437804\\
2.69256857886057	0.919086544055852\\
2.71639211689365	0.898023953086138\\
2.73974931981693	0.875394222359662\\
2.76256880980748	0.851086319707423\\
2.78477920904236	0.82498921296042\\
2.80630913969864	0.796991869949649\\
2.82708722395341	0.766983258506111\\
2.84704208398372	0.734852346460804\\
2.86610234196664	0.700488101644727\\
2.88419662007926	0.663779491888878\\
2.90125354049863	0.624615485024256\\
2.91720172540184	0.58288504888186\\
2.93196979696594	0.538477151292687\\
2.94548637736801	0.491280760087737\\
2.95768008878513	0.44118484309801\\
2.96847955339436	0.388078368154502\\
2.97781339337276	0.331850303088213\\
2.98561023089742	0.272389615730141\\
2.99179868814541	0.209585273911285\\
2.99630738729378	0.143326245462643\\
2.99906495051962	0.0735014982152164\\
3	0\\
};
\addplot [color=black, dashed, line width=1.0pt, forget plot]
  table[row sep=crcr]{0	0\\
0	1\\
1	1\\
1	2\\
2	2\\
2	1\\
3	1\\
3	0\\
};
\addplot [color=red, line width=2.0pt, draw=none, mark=o, mark options={solid, red}, forget plot]
  table[row sep=crcr]{0	0\\
0	1\\
1	1\\
1	2\\
2	2\\
2	1\\
3	1\\
3	0\\
};
\end{axis}
\end{tikzpicture}}
  \subcaption{Original NURBS curve.}
 \end{subfigure}
 \begin{subfigure}[c]{0.49\columnwidth}
  \resizebox{\textwidth}{!}{\definecolor{mycolor1}{rgb}{0.00000,0.44700,0.74100}\begin{tikzpicture}

\begin{axis}[width=4.519in,
height=3.564in,
at={(0.758in,0.481in)},
scale only axis,
axis lines=none,
xmin=-0.1,
xmax=3.1,
ymin=-0.1,
ymax=2.1,
axis background/.style={fill=white}
]

\addplot [loosely dotted, color=lightgray, line width=1.0pt, forget plot]
  table[row sep=crcr]{0	0\\
0.000935049480375794	0.073501498215216\\
0.00369261270621687	0.143326245462644\\
0.0082013118545938	0.209585273911285\\
0.0143897691025771	0.272389615730141\\
0.0221866066272373	0.331850303088213\\
0.0315204466056451	0.388078368154502\\
0.0423199112148708	0.44118484309801\\
0.0545136226319852	0.491280760087738\\
0.0680302030340587	0.538477151292687\\
0.0827982745981619	0.58288504888186\\
0.0987464595013653	0.624615485024256\\
0.115803379920739	0.663779491888878\\
0.133897658033355	0.700488101644727\\
0.152957916016283	0.734852346460804\\
0.172912776046592	0.766983258506111\\
0.193690860301355	0.796991869949649\\
0.215220790957642	0.824989212960419\\
0.237431190192522	0.851086319707423\\
0.260250680183067	0.875394222359662\\
0.283607883106348	0.898023953086138\\
0.307431421139434	0.919086544055852\\
0.331649916459396	0.938693027437804\\
0.356191991243305	0.956954435400998\\
0.380986267668232	0.973981800114433\\
0.405961367911246	0.989886153747112\\
0.431045914149419	1.00477852846803\\
0.456168528559821	1.0187699564462\\
0.481257833319523	1.03197146985062\\
0.506242450605595	1.04449410085029\\
0.531051002595107	1.0564488816142\\
0.555612111465131	1.06794684431137\\
0.579854399392736	1.07909902111079\\
0.603706488554994	1.09001644418146\\
0.627097001128975	1.10081014569239\\
0.649954559291749	1.11159115781258\\
0.672207785220387	1.12247051271103\\
0.69378530109196	1.13355924255673\\
0.714615729083538	1.1449683795187\\
0.734627691372191	1.15680895576593\\
0.753749958177883	1.16919181312656\\
0.77194268481368	1.18218744116617\\
0.789236050880565	1.19577629822302\\
0.805669710724612	1.20992666082024\\
0.821283318691891	1.22460680548099\\
0.836116529128475	1.2397850087284\\
0.850208996380436	1.25542954708561\\
0.863600374793845	1.27150869707578\\
0.876330318714775	1.28799073522203\\
0.888438482489297	1.30484393804752\\
0.899964520463482	1.32203658207538\\
0.910948086983405	1.33953694382876\\
0.921428836395135	1.3573132998308\\
0.931446423044746	1.37533392660465\\
0.941040501278308	1.39356710067344\\
0.950250725441894	1.41198109856032\\
0.959116749881576	1.43054419678842\\
0.967678228943426	1.44922467188091\\
0.975974816973515	1.4679908003609\\
0.984046168317916	1.48681085875156\\
0.9919319373227	1.50565312357601\\
0.99967177833394	1.52448587135741\\
1.00730534569771	1.54327737861889\\
1.01487229376007	1.5619959218836\\
1.02241227686711	1.58060977767468\\
1.02996494936489	1.59908722251528\\
1.03756996559948	1.61739653292852\\
1.04526697991697	1.63550598543757\\
1.05309564666341	1.65338385656556\\
1.06109562018488	1.67099842283562\\
1.06930655482745	1.68831796077092\\
1.0777681049372	1.70531074689458\\
1.0865199248602	1.72194505772975\\
1.09560166894251	1.73818916979957\\
1.10505299153021	1.75401135962719\\
1.11491354696938	1.76937990373574\\
1.12522298960608	1.78426307864838\\
1.13602097378638	1.79862916088824\\
1.14734715385637	1.81244642697846\\
1.1592411841621	1.82568315344219\\
1.17174195768621	1.83830795518632\\
1.18485876940742	1.85030260178615\\
1.19856246545049	1.86166595119652\\
1.21282132233857	1.87239800341742\\
1.2276036165948	1.88249875844886\\
1.24287762474232	1.89196821629083\\
1.25861162330427	1.90080637694334\\
1.2747738888038	1.90901324040639\\
1.29133269776405	1.91658880667997\\
1.30825632670816	1.92353307576408\\
1.32551305215928	1.92984604765873\\
1.34307115064054	1.93552772236391\\
1.36089889867509	1.94057809987963\\
1.37896457278607	1.94499718020589\\
1.39723644949663	1.94878496334268\\
1.4156828053299	1.95194144929\\
1.43427191680903	1.95446663804786\\
1.45297206045716	1.95636052961626\\
1.47175151279744	1.95762312399519\\
1.490578550353	1.95825442118465\\
1.509421449647	1.95825442118465\\
1.52824848720256	1.95762312399519\\
1.54702793954284	1.95636052961626\\
1.56572808319097	1.95446663804786\\
1.5843171946701	1.95194144929\\
1.60276355050337	1.94878496334268\\
1.62103542721393	1.94499718020589\\
1.63910110132491	1.94057809987963\\
1.65692884935946	1.93552772236391\\
1.67448694784072	1.92984604765873\\
1.69174367329184	1.92353307576408\\
1.70866730223595	1.91658880667997\\
1.7252261111962	1.90901324040639\\
1.74138837669573	1.90080637694334\\
1.75712237525768	1.89196821629083\\
1.7723963834052	1.88249875844886\\
1.78717867766143	1.87239800341742\\
1.80143753454951	1.86166595119652\\
1.81514123059258	1.85030260178615\\
1.82825804231379	1.83830795518632\\
1.8407588158379	1.82568315344219\\
1.85265284614363	1.81244642697846\\
1.86397902621362	1.79862916088824\\
1.87477701039392	1.78426307864838\\
1.88508645303062	1.76937990373574\\
1.89494700846979	1.75401135962719\\
1.90439833105749	1.73818916979957\\
1.9134800751398	1.72194505772975\\
1.9222318950628	1.70531074689458\\
1.93069344517255	1.68831796077092\\
1.93890437981512	1.67099842283562\\
1.94690435333659	1.65338385656556\\
1.95473302008303	1.63550598543757\\
1.96243003440052	1.61739653292853\\
1.97003505063511	1.59908722251528\\
1.97758772313289	1.58060977767468\\
1.98512770623993	1.5619959218836\\
1.99269465430229	1.54327737861889\\
2.00032822166606	1.52448587135741\\
2.0080680626773	1.50565312357601\\
2.01595383168208	1.48681085875156\\
2.02402518302648	1.4679908003609\\
2.03232177105657	1.44922467188091\\
2.04088325011842	1.43054419678843\\
2.04974927455811	1.41198109856032\\
2.05895949872169	1.39356710067344\\
2.06855357695525	1.37533392660465\\
2.07857116360486	1.3573132998308\\
2.0890519130166	1.33953694382876\\
2.10003547953652	1.32203658207538\\
2.1115615175107	1.30484393804752\\
2.12366968128523	1.28799073522203\\
2.13639962520615	1.27150869707578\\
2.14979100361956	1.25542954708561\\
2.16388347087152	1.2397850087284\\
2.17871668130811	1.22460680548099\\
2.19433028927539	1.20992666082024\\
2.21076394911943	1.19577629822302\\
2.22805731518632	1.18218744116617\\
2.24625004182212	1.16919181312656\\
2.26537230862781	1.15680895576593\\
2.28538427091646	1.1449683795187\\
2.30621469890804	1.13355924255673\\
2.32779221477961	1.12247051271103\\
2.35004544070825	1.11159115781258\\
2.37290299887103	1.10081014569239\\
2.39629351144501	1.09001644418146\\
2.42014560060726	1.07909902111079\\
2.44438788853487	1.06794684431137\\
2.46894899740489	1.0564488816142\\
2.4937575493944	1.04449410085029\\
2.51874216668048	1.03197146985062\\
2.54383147144018	1.0187699564462\\
2.56895408585058	1.00477852846804\\
2.59403863208875	0.989886153747112\\
2.61901373233177	0.973981800114433\\
2.64380800875669	0.956954435400998\\
2.6683500835406	0.938693027437804\\
2.69256857886057	0.919086544055852\\
2.71639211689365	0.898023953086138\\
2.73974931981693	0.875394222359662\\
2.76256880980748	0.851086319707423\\
2.78477920904236	0.82498921296042\\
2.80630913969864	0.796991869949649\\
2.82708722395341	0.766983258506111\\
2.84704208398372	0.734852346460804\\
2.86610234196664	0.700488101644727\\
2.88419662007926	0.663779491888878\\
2.90125354049863	0.624615485024256\\
2.91720172540184	0.58288504888186\\
2.93196979696594	0.538477151292687\\
2.94548637736801	0.491280760087737\\
2.95768008878513	0.44118484309801\\
2.96847955339436	0.388078368154502\\
2.97781339337276	0.331850303088213\\
2.98561023089742	0.272389615730141\\
2.99179868814541	0.209585273911285\\
2.99630738729378	0.143326245462643\\
2.99906495051962	0.0735014982152164\\
3	0\\
};
\addplot [color=lightgray, loosely dotted, line width=1.0pt, forget plot]
  table[row sep=crcr]{0	0\\
0	1\\
1	1\\
1	2\\
2	2\\
2	1\\
3	1\\
3	0\\
};
\addplot [color=lightgray, line width=1.0pt, draw=none, mark=o, mark options={solid, lightgray}, forget plot]
  table[row sep=crcr]{0	0\\
0	1\\
1	1\\
1	2\\
2	2\\
2	1\\
3	1\\
3	0\\
};

\addplot [color=mycolor1, line width=2.0pt, forget plot]
  table[row sep=crcr]{0	0\\
0.00140125240902796	0.0730352952865639\\
0.00552834456703947	0.141490513601821\\
0.012266278870426	0.205520306895453\\
0.0215000577155789	0.265279327117139\\
0.0331146834988897	0.32092222621656\\
0.0469951586167498	0.372603656143397\\
0.0630264854655507	0.42047826884733\\
0.0810936664416838	0.464700716278039\\
0.10108170394154	0.505425650385205\\
0.122875600361512	0.542807723118509\\
0.146360358097991	0.577001586427631\\
0.171420979547367	0.60816189226225\\
0.197942467106033	0.636443292572049\\
0.225809823170379	0.662000439306707\\
0.254908050136798	0.684987984415906\\
0.285122150401681	0.705560579849324\\
0.316337126361418	0.723872877556643\\
0.348437980412403	0.740079529487543\\
0.381309714951025	0.754335187591705\\
0.414837332373677	0.766794503818809\\
0.44890583507675	0.777612130118535\\
0.483400225456635	0.786942718440565\\
0.518205505909724	0.794940920734579\\
0.553206678832409	0.801761388950256\\
0.588288746621079	0.807558775037278\\
0.623336711672129	0.812487730945325\\
0.658235576381948	0.816702908624078\\
0.692870343146927	0.820358960023217\\
0.72712601436346	0.823610537092421\\
0.760887592427936	0.826612291781373\\
0.794040079736748	0.829518876039753\\
0.826468478686286	0.83248494181724\\
0.858057791672942	0.835665141063516\\
0.888693021093109	0.83921412572826\\
0.918259169343176	0.843286547761154\\
0.946641238819536	0.848037059111877\\
0.973724231918581	0.853620311730111\\
0.9993931510367	0.860190957565535\\
1.02353299857029	0.867903648567831\\
1.04602902012905	0.876912751175386\\
1.0668180225471	0.887312103432747\\
1.08595185256011	0.899060496543473\\
1.10349792255639	0.912098448988459\\
1.11952364492428	0.926366479248595\\
1.1340964320521	0.941805105804775\\
1.14728369632816	0.958354847137889\\
1.15915285014079	0.97595622172883\\
1.16977130587831	0.994549748058491\\
1.17920647592905	1.01407594460776\\
1.18752577268132	1.03447532985754\\
1.19479660852346	1.05568842228871\\
1.20108639584377	1.07765574038217\\
1.20646254703058	1.10031780261881\\
1.21099247447222	1.12361512747952\\
1.21474359055701	1.1474882334452\\
1.21778330767327	1.17187763899673\\
1.22017903820932	1.19672386261501\\
1.22199819455349	1.22196742278093\\
1.22330818909409	1.24754883797539\\
1.22417643421945	1.27340862667927\\
1.22467034231789	1.29948730737346\\
1.22485732577773	1.32572539853887\\
1.2248047969873	1.35206341865637\\
1.22458016833492	1.37844188620687\\
1.22425085220891	1.40480131967126\\
1.22388426099759	1.43108223753042\\
1.22354780708928	1.45722515826525\\
1.22330890287232	1.48317060035665\\
1.22323496073501	1.5088590822855\\
1.22339339306568	1.53423112253269\\
1.22385161225266	1.55922723957912\\
1.22467703068426	1.58378795190569\\
1.22593706074881	1.60785377799327\\
1.22769911483463	1.63136523632277\\
1.23003060533004	1.65426284537508\\
1.23299894462337	1.67648712363108\\
1.23667154510294	1.69797858957168\\
1.24111581915706	1.71867776167776\\
1.24639917917407	1.73852515843022\\
1.25258789549711	1.75746201737541\\
1.25970384146349	1.77545752973008\\
1.26771121712956	1.79251719951745\\
1.27657036814926	1.80864895760673\\
1.28624164017651	1.82386073486715\\
1.29668537886524	1.83816046216791\\
1.30786192986937	1.85155607037824\\
1.31973163884285	1.86405549036734\\
1.33225485143959	1.87566665300442\\
1.34539191331353	1.88639748915871\\
1.35910317011858	1.89625592969942\\
1.37334896750869	1.90524990549576\\
1.38808965113777	1.91338734741695\\
1.40328556665976	1.9206761863322\\
1.41889705972858	1.92712435311072\\
1.43488447599817	1.93273977862174\\
1.45120816112244	1.93753039373445\\
1.46782846075533	1.94150412931809\\
1.48470572055077	1.94466891624186\\
1.50180028616268	1.94703268537497\\
1.51907250324499	1.94860336758665\\
1.53648271745164	1.94938889374611\\
1.55399127443654	1.94939719472255\\
1.57155851985363	1.9486362013852\\
1.58914479935683	1.94711384460327\\
1.60671045860008	1.94483805524597\\
1.6242158432373	1.94181676418252\\
1.64162129892241	1.93805790228213\\
1.65888717130936	1.93356940041402\\
1.67597380605205	1.9283591894474\\
1.69284154880443	1.92243520025149\\
1.70945074522042	1.91580536369549\\
1.72576174095396	1.90847761064863\\
1.74173488165895	1.90045987198012\\
1.75733051298935	1.89176007855917\\
1.77250898059906	1.882386161255\\
1.78723063014203	1.87234605093682\\
1.80145580727218	1.86164767847385\\
1.81514485764344	1.8502989747353\\
1.82825812690973	1.83830787059038\\
1.8407588158379	1.82568315344219\\
1.85265284614363	1.81244642697846\\
1.86397902621362	1.79862916088824\\
1.87477701039392	1.78426307864838\\
1.88508645303062	1.76937990373574\\
1.89494700846979	1.75401135962719\\
1.90439833105749	1.73818916979957\\
1.9134800751398	1.72194505772975\\
1.9222318950628	1.70531074689458\\
1.93069344517255	1.68831796077092\\
1.93890437981512	1.67099842283562\\
1.94690435333659	1.65338385656556\\
1.95473302008303	1.63550598543757\\
1.96243003440052	1.61739653292853\\
1.97003505063511	1.59908722251528\\
1.97758772313289	1.58060977767468\\
1.98512770623993	1.5619959218836\\
1.99269465430229	1.54327737861889\\
2.00032822166606	1.52448587135741\\
2.0080680626773	1.50565312357601\\
2.01595383168208	1.48681085875156\\
2.02402518302648	1.4679908003609\\
2.03232177105657	1.44922467188091\\
2.04088325011842	1.43054419678843\\
2.04974927455811	1.41198109856032\\
2.05895949872169	1.39356710067344\\
2.06855357695525	1.37533392660465\\
2.07857116360486	1.3573132998308\\
2.0890519130166	1.33953694382876\\
2.10003547953652	1.32203658207538\\
2.1115615175107	1.30484393804752\\
2.12366968128523	1.28799073522203\\
2.13639962520615	1.27150869707578\\
2.14979100361956	1.25542954708561\\
2.16388347087152	1.2397850087284\\
2.17871668130811	1.22460680548099\\
2.19433028927539	1.20992666082024\\
2.21076394911943	1.19577629822302\\
2.22805731518632	1.18218744116617\\
2.24625004182212	1.16919181312656\\
2.26537230862781	1.15680895576593\\
2.28538427091646	1.1449683795187\\
2.30621469890804	1.13355924255673\\
2.32779221477961	1.12247051271103\\
2.35004544070825	1.11159115781258\\
2.37290299887103	1.10081014569239\\
2.39629351144501	1.09001644418146\\
2.42014560060726	1.07909902111079\\
2.44438788853487	1.06794684431137\\
2.46894899740489	1.0564488816142\\
2.4937575493944	1.04449410085029\\
2.51874216668048	1.03197146985062\\
2.54383147144018	1.0187699564462\\
2.56895408585058	1.00477852846804\\
2.59403863208875	0.989886153747112\\
2.61901373233177	0.973981800114433\\
2.64380800875669	0.956954435400998\\
2.6683500835406	0.938693027437804\\
2.69256857886057	0.919086544055852\\
2.71639211689365	0.898023953086138\\
2.73974931981693	0.875394222359662\\
2.76256880980748	0.851086319707423\\
2.78477920904236	0.82498921296042\\
2.80630913969864	0.796991869949649\\
2.82708722395341	0.766983258506111\\
2.84704208398372	0.734852346460804\\
2.86610234196664	0.700488101644727\\
2.88419662007926	0.663779491888878\\
2.90125354049863	0.624615485024256\\
2.91720172540184	0.58288504888186\\
2.93196979696594	0.538477151292687\\
2.94548637736801	0.491280760087737\\
2.95768008878513	0.44118484309801\\
2.96847955339436	0.388078368154502\\
2.97781339337276	0.331850303088213\\
2.98561023089742	0.272389615730141\\
2.99179868814541	0.209585273911285\\
2.99630738729378	0.143326245462643\\
2.99906495051962	0.0735014982152164\\
3	0\\
};
\addplot [color=black, dashed, line width=1.0pt, forget plot]
  table[row sep=crcr]{0	0\\
0	1\\
1.5	0.5\\
1	2\\
2	2\\
2	1\\
3	1\\
3	0\\
};
\addplot [color=red, line width=2.0pt, draw=none, mark=o, mark options={solid, red}, forget plot]
  table[row sep=crcr]{0	0\\
0	1\\
1.5	0.5\\
1	2\\
2	2\\
2	1\\
3	1\\
3	0\\
};
\end{axis}
\end{tikzpicture}}
  \subcaption{Modified NURBS curve where one control point is changed.}
 \end{subfigure}
 \caption{Visualization of the modification of the shape of a NURBS curve by moving one control point.}
 \label{fig:controlpoints}
\end{figure}
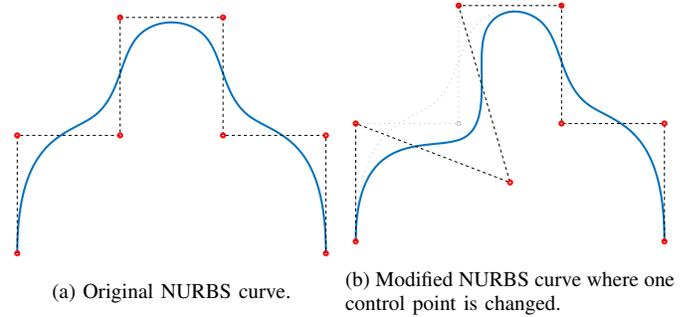

Isogeometric analysis is based on the idea of using freeform curves for an exact geometry representation of CAD models. This avoids errors in the geometry induced by mesh generation.
For this, the notion of \emph{B-splines}, i.e., piecewise polynomial functions, is of central importance.
B-splines are defined via \emph{knot vector} $\Xi=\left\{\xi_{1}, \xi_{2}, \dots, \xi_{n+p+1} \right\}$ where the knots $\xi_{i} \in [0,1]$ for all $i \in \{1,\dots,n+p+1\}$ are coordinates in the parametric space, $p$ is the degree of the B-spline and $n$ will be the dimension of the space. B-splines are defined by the recursive Cox-de Boor formula \cite{Piegl_1997aa}
\begin{align}
 B_{i,p}(\xi) = \frac{\xi-\xi_{i}}{\xi_{i+p}-\xi_{i}}B_{i,p-1}(\xi)
              + \frac{\xi_{i+p+1}-\xi}{\xi_{i+p+1}-\xi_{i+1}}B_{i+1,p-1}(\xi), 
\end{align}
for all $p \geq 1$ and for $p=0$ via
\begin{align}
 B_{i,0}(\xi) = \begin{cases}
            1 \quad \text{if } \xi_{i} \leq \xi < \xi_{i+1},\\
            0 \quad \text{else},
           \end{cases}
\end{align}
where $0/0=0$ is formally assumed.
\emph{NURBS} (Non-Uniform Rational B-splines) basis functions are then defined by 
\begin{align}
 N_{i,p}(\xi) = \frac{w_{i} B_{i,p}(\xi)}{\sum_{k=1}^{n} w_{k} B_{k,p}(\xi)},
\end{align}
where $w_{k}>0$ for all $k=1,2,\dots,n$ are so-called \emph{weights}. NURBS curves can then be defined by the NURBS basis functions and control points $\mathbf{P}_i$ as
\begin{align}
 \mathbf{C}(\xi) = \sum_{i=1}^{n}\mathbf{P}_i N_{i,p}(\xi).
\end{align}
A NURBS curve can be locally modified by moving the control points. This is visualized in Fig.~\ref{fig:controlpoints}.

\subsection{Isogeometric Analysis}\label{sec:iga}
\begin{figure}
\centering
 \begin{tikzpicture}
  \draw [fill=blue!10] (0,0) rectangle (1,1) node (ref) [pos = 0.5] {$\hat{{D}}$};
  \draw[->] (0,0)--(0,1.5);
  \draw[->] (0,0)--(1.5,0);
  \node (0) at (0,0) [anchor=north east] {$0$};
  \draw (1,2pt)--(1,-2pt) node [anchor=north] {$1$};
  \draw (2pt,1)--(-2pt,1) node [anchor=east] {$1$};
  \node () at (0,0) [anchor=north east] {$0$};
  
  \begin{scope}[shift={(5,0.5)}]
   \draw [fill=red!10] (0,0) node (middle){} circle (1.5);
   
\draw (-1.5,0)--(1.5,0);
   \draw (0,-1.5)--(0,1.5);
   
\node [anchor=north east] (p1) at (1,1) {${D}_1$};
   \node [anchor=north west] (p2) at (-1,1) {${D}_2$};
   \node [anchor=south west] (p3) at (-1,-1) {${D}_3$};
   \node [anchor=south east] (p4) at (1,-1) {${D}_4$};

   \draw [fill=white] (0,0) circle (0.5);
  \end{scope}
  
  \draw [->,teal, thick] (ref) to[in=110, out=30] node [pos=0.5, anchor=south east] {$\mathbf{F}_1(\hat{{D}})$} (p1);
  \draw [->,teal, thick] (ref) to[in=170, out=20] node [pos=0.5, anchor=north] {$\mathbf{F}_2(\hat{{D}})$} (p2);
  \draw [->,teal, thick] (ref) to[in=-170, out=-20]  node [pos=0.5, anchor=south] {$\mathbf{F}_3(\hat{{D}})$} (p3);
  \draw [->,teal, thick] (ref) to[in=-110, out=-30] node [pos=0.5, anchor=north east] {$\mathbf{F}_4(\hat{{D}})$} (p4);
 \end{tikzpicture}
 \vspace{-1em}
 \caption{Visualization of the mappings $\mathbf{F}_k$ from the reference domain $\hat{{D}}$ to the patches ${D}_k$ of a multipatch geometry.}
 \label{fig:multipatch}
\end{figure}
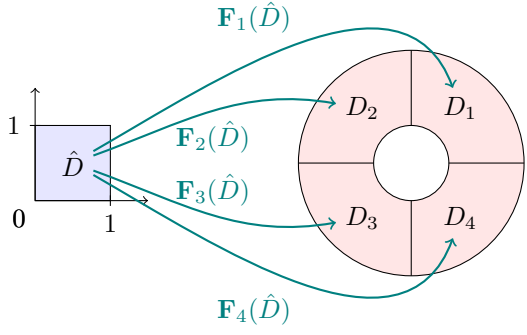

Using the same functions for the representation of the geometry as in CAD software, i.e., NURBS, has the advantage that there is no need for the construction of a finite element geometry, i.e., the mesh. 
In isogeometric analysis the geometry is represented by a smooth mapping 
\begin{align}
 \mathbf{F}: \hat{{D}}\rightarrow {D},
\end{align}
using NURBS as basis functions,
where $\hat{{D}}$ is the reference domain, i.e., in the 2D case the unit square, and ${D}$ is the physical domain. The mapping $\mathbf{F}$ can be directly obtained from CAD software. 
However, not all geometries can be represented by a regular transformation of the reference domain $\hat{{D}}=[0,1]^n$, e.g., domains with a hole. In this case, a multipatch approach can be used to represent the physical domain. The physical domain ${D}$ is subdivided into $k$ patches ${D}_k$ which can each be represented by a regular transformation of the reference domain ${D}_k=\mathbf{F}_k(\hat{{D}})$. A visualization is given in Fig.~\ref{fig:multipatch}. The patches have a consistent discretization with a one-to-one matching of the degrees of freedom at the interfaces as in classical FEM, leading to a $C^0$ smoothness of the solution at the patch interfaces.
Thus, the geometry of the CAD models can be exactly represented by a multipatch model \cite{Buffa_2015aa}
and there is no need for the generation of a computational mesh that approximates the geometry, e.g., a triangulation. For the analysis, IGA uses the same setting as the classical finite element analysis with the exception of using B-splines or NURBS as basis and test functions $w_j$ for $j=1,\dots , N_q$. 

\section{Numerical Shape Optimization} \label{sec:num_shape_opti}
We solve the optimization problem \eqref{eq_minTHD}--\eqref{pde_constraint_weak} by means of a gradient-based shape optimization algorithm based on the shape derivative \eqref{eq_shape_der}. By solving an auxiliary boundary value problem of the form
\begin{align} \label{eq_aux_prob}
	b(\mathbf{W}, \mathbf{Z}) = d \mathcal{J}(\Omega; \mathbf{Z}) \quad \forall \mathbf{Z}
\end{align}
with some positive definite bilinear form $b(\cdot, \cdot)$ satisfying $b(\mathbf{Z},\mathbf{Z})>0$ for all vector fields $\mathbf{Z}$, we can extract a shape gradient $\mathbf{W}$, which satisfies $d \mathcal{J}(\Omega; \mathbf{W}) = b(\mathbf{W},\mathbf{W}) >0$. Thus, moving the control points of the motor geometry a small distance $\delta$ into the direction of $\mathbf{W}$ will yield an increase of the objective function $\mathcal{J}$. Likewise, since $d\mathcal{J}(\Omega;\mathbf{W})$ is linear in $\mathbf{W}$, a decrease can be achieved by moving the control points into the direction of the negative shape gradient $-\mathbf{W}$. The auxiliary boundary value problem \eqref{eq_aux_prob} can be interpreted as finding a Riesz representative $\mathbf{W}$ of the functional $d \mathcal{J}(\Omega, \cdot)$ with respect to the metric given by $b(\cdot, \cdot)$. Of course, here different bilinear forms $b(\cdot, \cdot)$ can be chosen which amount to shape gradients in different metrics. In our algorithm, we choose
\begin{align} \label{eq_b_W_Z}
	b(\mathbf{W}, \mathbf{Z} ) = \int_D \mathbf{DW} : \mathbf{DZ} + \mathbf{W} \cdot \mathbf{Z} \; {\mathrm{d}} x
\end{align}
where we used $\mathbf{A}:\mathbf{B} = \sum_{i,j=1}^n A_{ij} B_{ij}$ denotes the Frobenius inner product for two matrices $\mathbf{A}, \mathbf{B} \in \mathbb R^n$ and $\mathbf{DW}$, $\mathbf{DZ}$ denote the Jacobi matrices of $\mathbf{W}$, $\mathbf{Z}$, respectively. Our algorithm consists in iteratively determining a descent vector field $\mathbf{W}$ by solving the auxiliary boundary value problem \eqref{eq_aux_prob} with the bilinear form $b(\cdot, \cdot)$ given by \eqref{eq_b_W_Z} and moving the domain a distance $\delta$ into this direction. The step size $\delta$ is chosen as the maximum of the set $\{1, 1/2, 1/4 , \dots \}$ such that, no intersections of the patches occur and the objective value is decreased. When no further improvement can be achieved, the algorithm terminates. An overview of the shape optimization algorithm is given in Fig.~\ref{fig:shape_opti_overview}.
\tikzstyle{decision} = [diamond, aspect=4, draw, fill=green!7, text badly centered, inner sep=0pt]
    \tikzstyle{block} = [rectangle, draw, fill=green!3, 
    text width=0.5\columnwidth, text centered, rounded corners, minimum height=2em]
    \tikzstyle{line} = [draw, -latex']
 \begin{figure}
  \begin{center}
  \resizebox{\columnwidth}{!}{
    \begin{tikzpicture}[every node/.style={font=\footnotesize}]
\node [block] (goal) {choose functional $\mathcal{J}(\Omega)$ to minimize, initial geometry $\Omega_0$};
        \node [node distance = 4em, block, below of=goal] (shapederiv) {find shape derivative $d \mathcal{J}(\Omega_i \mathchar\numexpr"6000+`;\relax \mathbf{W})$};
        \node [node distance = 4em, block, below of=shapederiv] (descent) {find descent direction $\mathbf{W}_i$ by solving $$b(\mathbf{W}_i, \mathbf{Z}) = d \mathcal{J}(\Omega_i \mathchar\numexpr"6000+`;\relax \mathbf{Z}) \quad \forall \mathbf{Z}$$};
        \node [node distance = 6em, block, below of=descent] (transform) {move every point $\x$ of $\Omega_i$ a small distance $\delta_i$ in direction of $-\mathbf{W}_i$ to get $$\Omega_{i+1} = (id - \delta_i \mathbf{W}_i)(\Omega_i)$$ where $\delta_i$ such that $\mathcal{J}(\Omega_{i+1})<\mathcal{J}(\Omega_i)$};
        \node [decision,xshift=4cm, right of=transform] (decide) {$\mathcal{J}(\Omega_{i})-\mathcal{J}(\Omega_{i+1})<\mathrm{tol}$};
        \node [node distance = 5em, block, below of=decide, text width=0.3\columnwidth] (terminate) {terminate};
        
        \path [line] (goal) -- (shapederiv) node
        [anchor = west, pos = 0.5] {$i = 0$};
        \path [line] (shapederiv) -- (descent);
        \path [line] (descent) -- (transform);
        \path [line] (transform) -- (decide);
        \path [line] (decide) -- (terminate) node
        [anchor = west, pos = 0.5] {yes};
        \path [line] (decide) |- (shapederiv) node
        [anchor = west, pos = 0.25] {no} node
        [anchor = south, pos = 0.75] {$i=i+1$};
    
    \end{tikzpicture}
    }
    \end{center}
    \caption{Overview of the shape optimization algorithm.}\label{fig:shape_opti_overview}
  \end{figure}
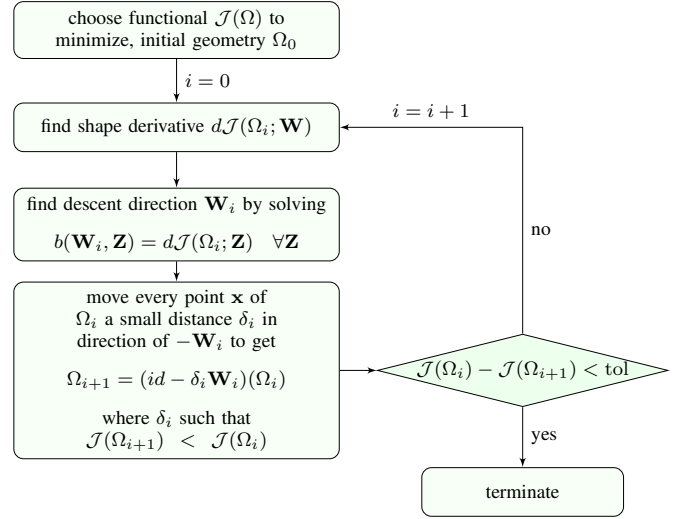

\section{Results of the Optimization} \label{sec:results}
\begin{figure}
 \center
 \begin{subfigure}[c]{0.49\columnwidth}
  \center
  \begin{tikzpicture}
    \begin{scope}
        \node[inner sep=0pt] at (0,0)
            {\includegraphics[width=\textwidth,clip]{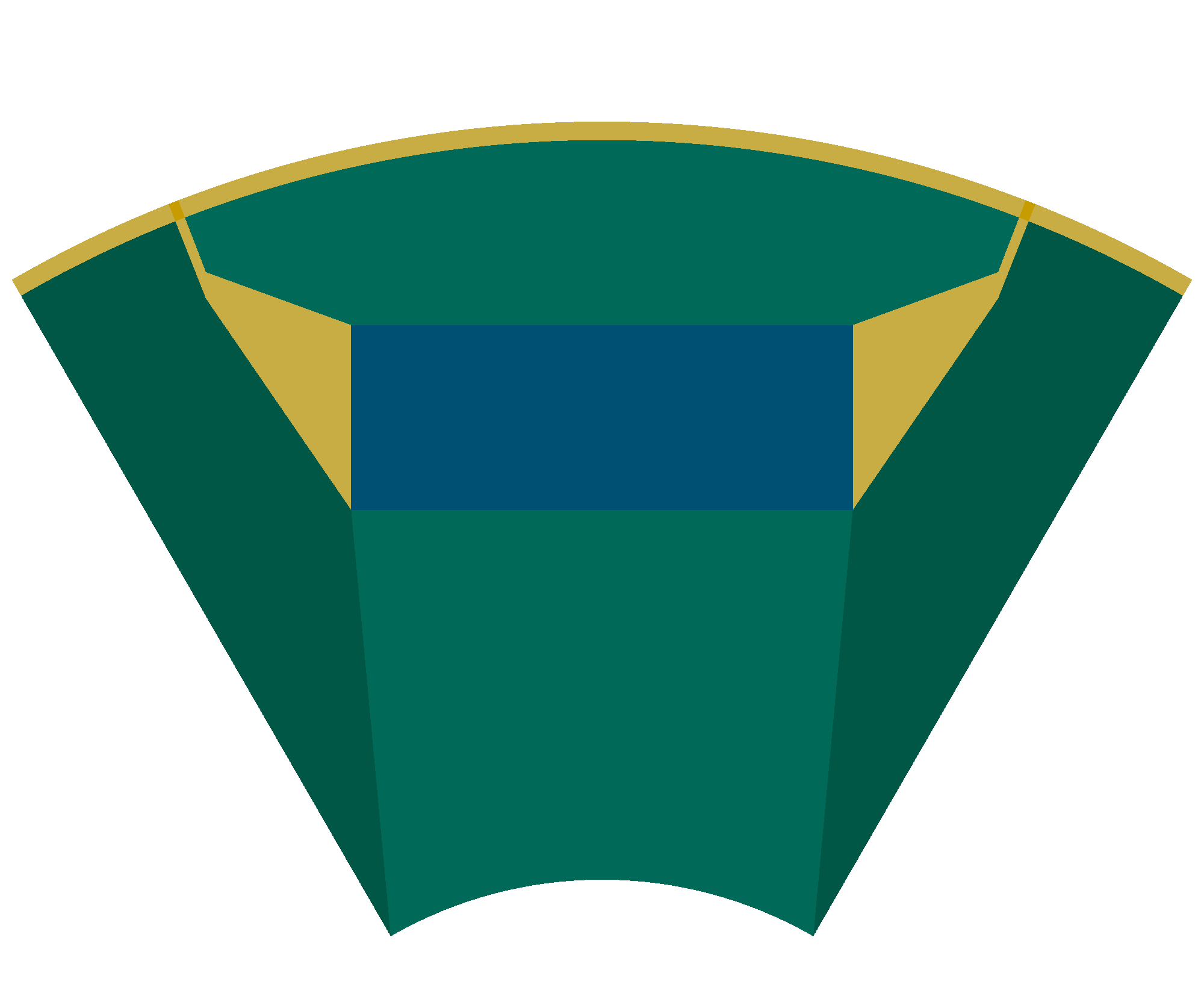}};
\end{scope}
  \end{tikzpicture}
  \subcaption{Original design of the machine.\label{fig:pmsm_nicht_opti}}
 \end{subfigure}
 \begin{subfigure}[c]{0.49\columnwidth}
  \center
  \begin{tikzpicture}
            \begin{scope}
                \node[inner sep=0pt] at (0,0)
                {\includegraphics[width=\textwidth]{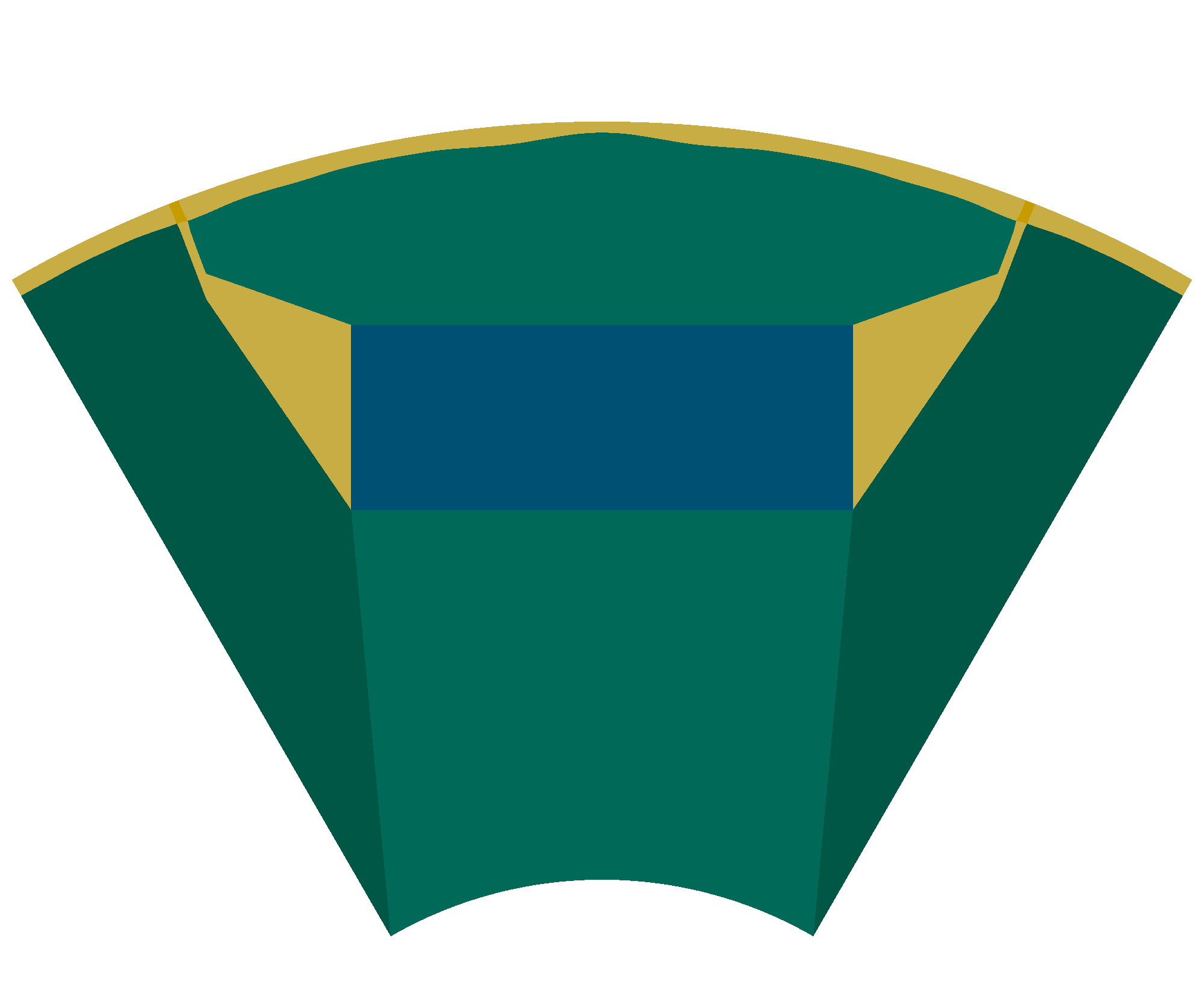}};
                {
                \node[inner sep=0pt, opacity=0.5] at (0,0)
                {\includegraphics[width=\textwidth]{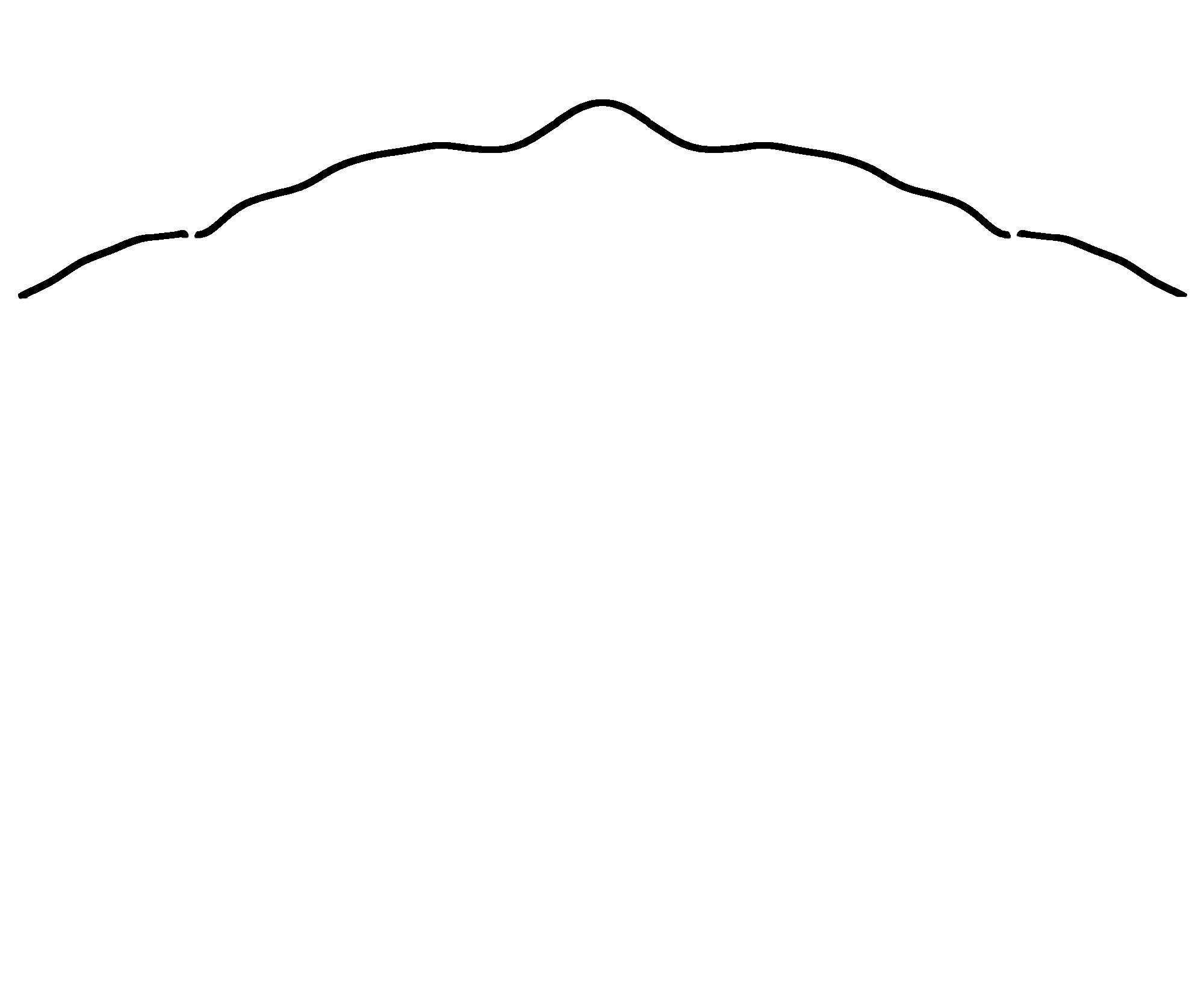}};
                }
            \end{scope}
            \end{tikzpicture}
\subcaption{Optimized design of the machine.\label{fig:pmsm_opti}}
 \end{subfigure}
\caption{Results of shape optimization \eqref{eq_minTHD} in generator mode under no load condition. The shape of the rotor was optimized, minimizing the total harmonic distortion. Shape and position of the permanent magnet was fixed. Figure (a) shows the original design, figure (b) shows the optimized design. The black line shows the changes in the rotor amplified by a factor of 5. \label{fig:rotor_design_opt}}
 \vspace{-0.5em}
\end{figure}

The method described in Section \ref{sec:num_shape_opti} is used to minimize the total harmonic distortion of the electromotive force by shape optimizing the rotor of a 6-pole permanent magnet synchronous machine (for the description of geometry and material coefficients see \cite[Chapter V.A]{Bontinck_2018af} and to download the geometry of the machine as an IGES file see \cite{Merkel_2020git}) in generator mode under no load condition. 
The shape and position of the permanent magnets in the rotor are fixed. 
The machine is discretized by $N_{\mathrm{DoF}}=4354$ degrees of freedom using B-splines of degree $p=2$ as basis functions and linearized material laws. The implementation is based on GeoPDEs \cite{Falco_2011aa}. The level of refinement is chosen such that the relative error in the $L^2$-norm of the magnetic vector potential $u$ in the domain is smaller than $\si{10^{-3}}$.
The coupling of stator and rotor domain is realized using ${N_\Gamma} = 36$ harmonic basis functions.  The simulation of a rotation of $\SI{120}{\degree}$ with $N_{\alpha}=120$ takes about $\SI{10}{\second}$. The optimization process takes about $\SI{2}{h}$ and is carried out in Matlab\textsuperscript{\textregistered} R2019a on a 6-core machine (Intel\textsuperscript{\textregistered}
 Core\texttrademark{} i7-5820K CPU) with \SI{16}{GB} RAM. The original and optimized design of the rotor can be seen in Fig.~\ref{fig:rotor_design_opt}. The total harmonic distortion is reduced by more than $75\,\%$ in 59 iterations. The total harmonic distortion in the iterations of the shape optimization algorithm and the Fourier coefficients of the original design compared to the Fourier coefficients of the optimized design can be seen in \cref{fig:THD_opt,fig:fourier_coef_opt}.

     \begin{figure}
     \centering
        \begin{tikzpicture}
           \begin{axis}[ylabel={$\thd({\mathcal E})$},xlabel={iteration}, width = \columnwidth, height=0.5\columnwidth, ylabel near ticks, xlabel near ticks,tick label style={font=\footnotesize}, label style={font=\footnotesize}, legend style={font=\footnotesize},xtick={0,10,20,30,40,50,60}, ytick={0.02,0.04,0.06,0.08,0.1}, yticklabel style={/pgf/number format/fixed,/pgf/number format/precision=2},scaled y ticks=false]
                \addplot  [blue,  thick
                ] table [x index=0, y index=1, col sep=comma] {images/thd_iter1-60.csv};
                \addplot  [red,mark=diamond*,only marks, mark size = 1.5pt,  thick] table [x index=0, y index=1, col sep=comma] {images/thd_jmag.csv};
                \addplot  [green,mark=*,only marks, mark size = 1.5pt,  thick] table [x index=0, y index=1, col sep=comma] {images/thd_jmag_nonlinear.csv};
                \legend{{IGA linear}, {JMAG linear}, {JMAG nonlinear}}
            \end{axis}
        \end{tikzpicture}
        \caption{Total harmonic distortion of the electromotive force ${\mathcal E}$ for each iteration of the shape optimization \eqref{eq_minTHD} in generator mode under no load condition computed with IGA and with JMAG\textsuperscript{\textregistered} as a reference. \label{fig:THD_opt}}
     \end{figure}
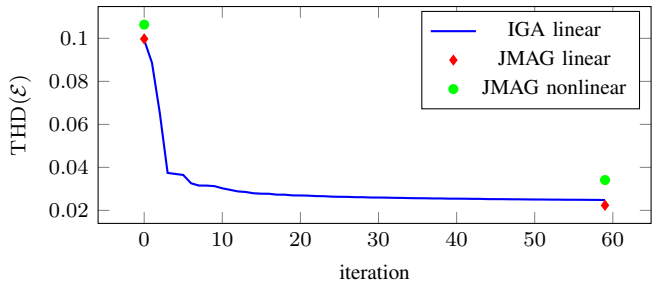
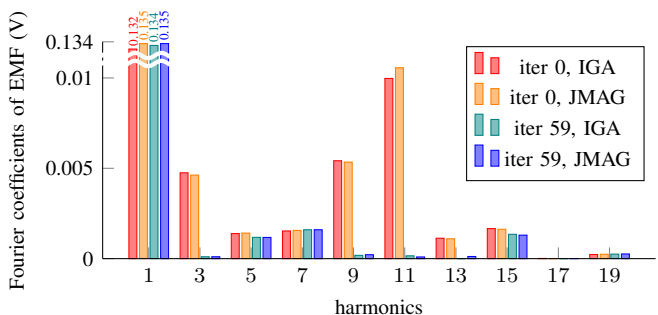
\begin{figure}
     \centering
        \begin{tikzpicture}
            \begin{axis}[
            width=\columnwidth,
            height = 0.5\columnwidth,
            height = 0.5\columnwidth,
		  ytick={0, 0.005, 0.01, 0.012},
		  yticklabels={0, 0.005, 0.01, 0.134},
      yticklabel style={
	      /pgf/number format/fixed,
	      /pgf/number format/precision=4
      },
      scaled y ticks=false,
      ybar=1pt,
	  bar width=3pt,
	  ymin=0,
	  axis on top,
	  xlabel={harmonics},ylabel={Fourier coefficients of EMF (V)}, ylabel near ticks, xlabel near ticks, ylabel near ticks,
	  xtick align=inside,
	  xtick=data,
	  ymax=1.2e-2,
	  restrict y to domain*=0:1.2e-2, visualization depends on=rawy\as\rawy, after end axis/.code={ \draw [ultra thick, white, decoration={snake, amplitude=1pt}, decorate] (rel axis cs:-0.1,0.9) -- (rel axis cs:1,0.9);
		  \draw [yshift=3pt,ultra thick, white, decoration={snake, amplitude=1pt}, decorate] (rel axis cs:-0.1,0.9) -- (rel axis cs:1,0.9);
	      },
	  axis lines*=left,
	  clip=false,
	  tick label style={font=\footnotesize}, 
	  label style={font=\footnotesize}, legend style={font=\footnotesize}
	  ]
		      \addplot[red, fill=red!50] table [restrict y to domain*=0:1.16e-2,x index=0, y index=1, col sep=comma] {images/harmonics_fourierCoefficients_iter1-60.csv};
		      \addplot[orange, fill=orange!50] table [restrict y to domain*=0:1.19e-2,x index=0, y index=1, col sep=comma] {images/harmonics_fouriercoeffs_jmag_iter0.csv};
		      \addplot[teal, fill=teal!50] table [restrict y to domain*=0:1.18e-2,x index=0, y index=60, col sep=comma] {images/harmonics_fourierCoefficients_iter1-60.csv};
		      \addplot[blue, fill=blue!50] table [restrict y to domain*=0:1.19e-2,x index=0, y index=1, col sep=comma] {images/harmonics_fouriercoeffs_jmag_iter59.csv};
		        \legend{{iter 0, IGA}, {iter 0, JMAG}, {iter 59, IGA}, {iter 59, JMAG}}
		        \node [red] (0) at (axis cs:0.4,0.0126) [rotate=90] {{\fontsize{5}{6}\selectfont 0.132}};
		        \node [orange] (0) at (axis cs:0.8,0.0129) [rotate=90] {{\fontsize{5}{6}\selectfont 0.135}};
		        \node [teal] (0) at (axis cs:1.2,0.0128) [rotate=90] {{\fontsize{5}{6}\selectfont 0.134}};
		        \node [blue] (0) at (axis cs:1.6,0.0129) [rotate=90] {{\fontsize{5}{6}\selectfont 0.135}};
		  \end{axis}
        \end{tikzpicture}
        \caption{Fourier coefficients of the electromotive force for the original design and for the optimized design after 59 iterations. Computed using linear material laws. \label{fig:fourier_coef_opt}}
    \end{figure}
\begin{table}
 \centering
 \begin{tabular}{ |c|c|c|  }
  \hline
      & $\thd({\mathcal E})$  & $\thd({\mathcal E})$ \\
      & original design & optimized design \\
  \hline
  IGA linear & 0.099268 & 0.024793 \\
  JMAG linear & 0.099754  & 0.02234 \\
  JMAG nonlinear & 0.10639 & 0.034106 \\
\hline
\end{tabular}
\caption{Total harmonic distortion of the electromotive force ${\mathcal E}$ for original and the optimized rotor design computed with IGA and with JMAG\textsuperscript{\textregistered} for linear and nonlinear material laws.}
\label{table:thd}
\end{table}
\section{Validation}
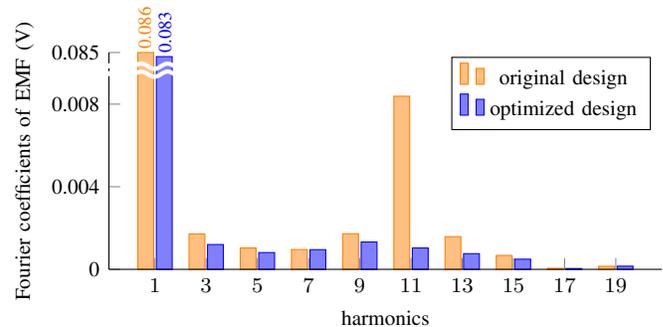
\begin{figure}[]
        \begin{tikzpicture}
            \begin{axis}[
            width=\columnwidth,
            height = 0.5\columnwidth,
            height = 0.5\columnwidth,
		  ytick={0, 0.004, 0.008, 0.0105},
		  yticklabels={0, 0.004, 0.008, 0.085},
      yticklabel style={
	      /pgf/number format/fixed,
	      /pgf/number format/precision=4
      },
      scaled y ticks=false,
      ybar=1pt,
	  bar width=6pt,
	  ymin=0,
	  axis on top,
	  xlabel={harmonics},ylabel={Fourier coefficients of EMF (V)}, ylabel near ticks, xlabel near ticks, ylabel near ticks,
	  xtick align=inside,
	  xtick=data,
	  ymax=1.05e-2,
	  restrict y to domain*=0:1.05e-2, visualization depends on=rawy\as\rawy, after end axis/.code={ \draw [ultra thick, white, decoration={snake, amplitude=1pt}, decorate] (rel axis cs:-0.1,0.9) -- (rel axis cs:1,0.9);
		  \draw [yshift=3pt,ultra thick, white, decoration={snake, amplitude=1pt}, decorate] (rel axis cs:-0.1,0.9) -- (rel axis cs:1,0.9);
	      },
	  axis lines*=left,
	  clip=false,
	  tick label style={font=\footnotesize}, 
	  label style={font=\footnotesize}, legend style={font=\footnotesize}
	  ]
		      \addplot[orange, fill=orange!50] table [restrict y to domain*=0:1.05e-2,x index=0, y index=1, col sep=comma] {images/harmonics_fouriercoeffs_jmag_iter0_nonlinear.csv};
		      \addplot[blue, fill=blue!50] table [restrict y to domain*=0:1.03e-2,x index=0, y index=1, col sep=comma] {images/harmonics_fouriercoeffs_jmag_iter59_nonlinear.csv};
		        \legend{{original design}, {optimized design}}
		        \node [orange] (0) at (axis cs:0.6,0.0117) [rotate=90] {{\fontsize{7}{6}\selectfont 0.086}};
		        \node [blue] (0) at (axis cs:1.4,0.0115) [rotate=90] {{\fontsize{7}{6}\selectfont 0.083}};
		  \end{axis}
        \end{tikzpicture}
        \caption{Fourier coefficients of the electromotive force for the original design and for the optimized design after 59 iterations considering nonlinear material laws in stator and rotor core (i.e. electrical steel M530-50A), calculated using JMAG\textsuperscript{\textregistered}.}\label{fig:fourier_coef_nonlinear}
    \end{figure}

Exploiting the geometry representation which is given directly in NURBS, i.e., the representation commonly used in computer-aided design (CAD), the model can be easily imported into computer-aided-engineering (CAE) software, e.g., using the IGES (Initial Graphics Exchange Specification) format. The results of the optimization in \cref{sec:results} is validated by using JMAG\textsuperscript{\textregistered} as reference solution. The JMAG\textsuperscript{\textregistered} model has 281198 elements of order $1$ and the meshing takes about $\SI{15}{\second}$. The simulation of a single rotation of $\SI{120}{\degree}$ with $N_{\alpha}=120$ takes about $\SI{30}{\minute}$. The results of the JMAG\textsuperscript{\textregistered} simulation using linearized material laws is shown in \cref{fig:THD_opt,fig:fourier_coef_opt}.
They are in very good agreement with the results from GeoPDEs for both geometries.
To investigate the impact of saturation on the results, the material of the rotor is modeled in JMAG\textsuperscript{\textregistered} by the nonlinear material curve M530-50A for both, the original and the optimized geometry. The results for the THD of the EMF can be seen in Fig.~\ref{fig:THD_opt} and \cref{table:thd} where still a reduction of $68\,\%$ is achieved. 
The comparison of the Fourier coefficients of the electromotive force for the original and the optimized machine considering nonlinear material laws is depicted in Fig.~\ref{fig:fourier_coef_nonlinear}.
In conclusion, the consideration of fully nonlinear material laws increases the THD but the optimization using the linear model is still very effective.
\section{Conclusion}
This paper has proposed a freeform shape optimization methodology based on isogeometric analysis. The optimized geometry exhibits a significantly reduced total harmonic distortion, i.e., by a factor of four. It has been demonstrated that the spline-based geometry representation allows easy integration into existing workflows using proprietary software.

\section*{Acknowledgements}
This work is supported by the German BMBF by the PASIROM project (grant nr. 05M2018RDA), the `Excellence Initiative' of the German Federal and State Governments and by the Graduate School of Computational Engineering at Technische Universit\"at Darmstadt. The authors thank C. Mellak, H. De Gersem, Z. Bontinck and J. Corno for their help and the fruitful discussions.

\end{document}